\documentclass[aps,prx,twocolumn,amsmath,amssymb,floatfix,superscriptaddress]{revtex4-2} %rmp

\usepackage[outline]{contour}

\usepackage{color}
\usepackage{bbm}
\usepackage{graphicx}
\usepackage{dcolumn}
\usepackage[utf8]{inputenc}
\usepackage{graphicx}
\usepackage{enumitem}
\usepackage{epstopdf}
\usepackage{amsthm}
\usepackage{amsmath}
\usepackage{empheq}
\usepackage{bbm}
\usepackage{braket}
\usepackage{amssymb}
\usepackage{mathtools}
\usepackage[usenames,dvipsnames]{xcolor}
\usepackage{latexsym}
\usepackage[colorlinks=true,citecolor=Cerulean,linkcolor=RubineRed,urlcolor=Cerulean,hypertexnames=false]{hyperref}
\usepackage[capitalize]{cleveref}
\usepackage{subcaption}

\usepackage{ragged2e} % for the \justifying macro
\DeclareCaptionJustification{justified}{\justifying}

\newcommand{\cov}{\textnormal{cov}}
\newcommand{\var}{\textnormal{var}}

\newcommand{\id}{\mathbbm{I}}
\newcommand{\tr}[1]{\operatorname{\textnormal{Tr}}\left( {#1} \right)}

\newcommand{\dist}{\mathcal{D}_{\textnormal{B}}}

\newcommand{\robust}{\mathcal{F}}

\newcommand{\length}{\mathcal{L}}

\newcommand{\vect}[1]{\boldsymbol{#1}} 
\newcommand{\at}{\Big\vert_{\vect{\delta \theta = 0}}}

\newcommand{\ketbra}[2]{\ensuremath{|#1\rangle\!\langle#2|}}

\begin{document}
%\clearpage
%\pagenumbering{arabic}
%\normalsize

\title{
Resilience--Runtime Tradeoff Relations for Quantum Algorithms 
%1 Identifying Noise-Resilient Compilations of Quantum Algorithms
%\\
%geometric bounds on the noise resilience of quantum algorithms\\
}

\date{\today}

\author{Luis Pedro Garc\'ia-Pintos}
\email{lpgp@lanl.gov}
\affiliation{Theoretical Division (T4), Los Alamos National Laboratory, Los Alamos, New Mexico 87545, USA}

\author{Tom O'Leary}
\affiliation{Theoretical Division (T4), Los Alamos National Laboratory, Los Alamos, New Mexico 87545, USA}
\affiliation{Clarendon Laboratory, University of Oxford, Parks Road, Oxford OX1 3PU, United Kingdom}

\author{Tanmoy Biswas}
\affiliation{Theoretical Division (T4), Los Alamos National Laboratory, Los Alamos, New Mexico 87545, USA}

\author{Jacob Bringewatt}
\affiliation{Joint Center for Quantum Information and Computer Science,  
University of Maryland, College Park, Maryland 20742, USA} 
\affiliation{Joint Quantum Institute, University of Maryland, College Park, Maryland 20742, USA} 

\author{Lukasz Cincio}
\affiliation{Theoretical Division (T4), Los Alamos National Laboratory, Los Alamos, New Mexico 87545, USA}

\author{Lucas T. Brady}
\affiliation{Quantum Artificial Intelligence Laboratory, NASA Ames Research Center, Moffett Field, California 94035, USA}

\author{Yi-Kai Liu}
\affiliation{Joint Center for Quantum Information and Computer Science,  
University of Maryland, College Park, Maryland 20742, USA} 
\affiliation{Applied and Computational Mathematics Division, National Institute of Standards and Technology, Gaithersburg, MD 20899, USA}

\begin{abstract}

A leading approach to algorithm design aims to minimize the number of operations in an algorithm's compilation. One intuitively expects that reducing the number of operations may decrease the chance of errors. This paradigm is particularly prevalent in quantum computing, where gates are hard to implement and noise rapidly decreases a quantum computer's potential to outperform classical computers. 
Here, we find that minimizing the number of operations in a quantum algorithm can be counterproductive, leading to a noise sensitivity that induces errors when running the algorithm in non-ideal conditions. To show this, we develop a framework to characterize the resilience of an algorithm to perturbative noises (including coherent errors, dephasing, and depolarizing noise). Some compilations of an algorithm can be resilient against certain noise sources while being unstable against other noises. 
We condense these results into a tradeoff relation between an algorithm's number of operations and its noise resilience. 
We also show how this framework can be leveraged to identify compilations of an algorithm that are better suited to withstand certain noises.

\end{abstract}

\maketitle

%Some natural phenomena, including certain chemical reactions, high-energy physics processes, and dynamics of many-body quantum systems, are intractable to simulate by classical computing devices. As long as no efficient classical algorithm exists, an alternative is to build a quantum device to simulate these processes~\cite{manin1980computable,benioff1980computer,Feynman82}. 
The realization that a quantum device may be able to efficiently simulate processes and solve computational tasks that no classical computer can has seeded decades of research in quantum computing~\cite{preskill2023quantum}. 
%Research in quantum computing 
Such research can be coarsely divided into two lines:
\begin{itemize}
 \item[(a)] designing quantum algorithms with potential practical advantages over classical algorithms~\cite{montanaro2016quantum}, and 
\item[(b)] physically building a quantum computer~\cite{ladd2010quantum}, including developing techniques to overcome the crippling effect of noise on quantum systems~\cite{lidar2013quantum, RoffeQEC2019, RevModPhysErrorMitigation, Klimov2024}. 
\end{itemize} 

At these early stages in the process of building a useful quantum computer, research directions (a) and (b) often evolve independently. 
For instance, theoretical work on quantum algorithms involves the technically challenging task of proving a quantum advantage over their best-known classical counterparts. Typically, this work does not incorporate details of current experimental devices in the algorithm's design.
Meanwhile, realizable devices have not reached sizes and coherence times that allow running an algorithm that can empirically demonstrate useful quantum advantage. Thus, %experimental, theoretical, and engineering
work along line (b) is more preoccupied with constructing scalable coherent devices, implementing larger quantum circuits, or providing proof-of-principle demonstrations with small numbers of logical qubits~\cite{Sivak2023,google2023suppressing,bluvstein2024logical,da2024demonstration} than with the details of algorithmic design.

Still, recent advances suggest we may be fast approaching a regime where merging both research lines %(a) and (b) 
will become important as we seek to implement useful examples of quantum computing.  %In particular, identifying algorithms suitable to run on a computing platform affected by a given noise is a problem that straddles both research lines. 
%Addressing this problem is this article's main motivation. 
Some quantum computing platforms have reached a level of sophistication where they have an overcomplete set of gates, i.e., they can synthesize a desired unitary transformation in many different ways. This has inspired much work on compilers for finding the most efficient way to implement a given quantum algorithm. 

However, one can consider a different possibility: compiling a quantum algorithm to optimize its resilience to noise (without error correction). 
This is a rich kind of optimization because, as we will show, a smaller circuit is not always more noise-resilient. 
Performing this optimization effectively is a well-recognized challenge for quantum compilers~\cite{chong2017programming}. 
%This approach 
A noise-optimized approach to compiling crucially exploits the fact that a given type of noise affects different quantum circuits differently. 
This paper aims to present a theoretical framework for analyzing these effects.

Our main contribution consists of 
%We develop
a broad framework to discriminate algorithms' resilience to the noise affecting a computing device. 
%xxxxx
%We do so by deriving
In our framework, easy-to-evaluate metrics 
%that 
quantify the effect of perturbative noise on quantum computations (Sec.~\ref{sec:robustness}). %By quantifying the impact of noise, we can study an algorithm's resilience to it. 
%In our perturbative framework, 
Noise resilience can be evaluated in terms of the ideal (unperturbed) dynamics of the computer, thereby side-stepping expensive simulations of noisy dynamics.

%We develop a broad framework to discriminate algorithms' resilience to the noise affecting a computing device. We do so by deriving easy-to-evaluate metrics that quantify the effect of perturbative noise on quantum computations (Sec.~\ref{sec:robustness}). In our perturbative framework, noise resilience can be evaluated in terms of the ideal (unperturbed) dynamics of the computer, thereby side-stepping expensive simulations of noisy dynamics.

We find that an algorithm can be resilient to one noise process while being fragile to other noises. At the same time, different compilations of an algorithm show different noise resilience (Sec.~\ref{sec:robustness-incoherent}). This realization leads to the notion of noise-tailored compilations of a quantum algorithm. We illustrate our results by characterizing the resilience to different noise sources of two error-detection circuits, of two-qubit gates affected by correlated errors, and of optimized vs. adiabatic annealing algorithms.

Since the noise affecting a computing device is platform-dependent, the suitability of an algorithm's compilation will also be platform-dependent. 
We show that the most resilient compilations do not, in general, correspond to the ones with shorter runtimes/gates. 
Finally, we prove an inequality that constrains an algorithm's runtime (or number of gates for digital circuits) and its noise resilience (Sec.~\ref{sec:tradeoffs}). The inequality sets a minimum on an algorithm's runtime/number of gates needed to achieve a desired noise-resilience.

\section{Noise resilience of quantum algorithms}
\label{sec:robustness}
Consider the implementation of a quantum algorithm in ideal conditions. The computer's state evolves by
\begin{align}
\label{eq:idealdynamics}
    \ket{\psi_0} \xrightarrow[\textcolor{blue}{\textnormal{(ideal)}}]{\textnormal{layer $1$}}  \ket{\psi_1} \cdots \xrightarrow[\textcolor{blue}{\textnormal{(ideal)}}]{\textnormal{layer $l$}} \ket{\psi_l} \cdots \xrightarrow[\textcolor{blue}{\textnormal{(ideal)}}]{\textnormal{layer $D$}}  \ket{\psi_D}.
\end{align}
The final state $\ket{\psi_D} \coloneqq \prod_{l=1}^D \bigotimes_{q = 1}^{\mathcal{N}_l} M_l^q V_l^q \ket{\psi_0}$ is obtained after $\ket{\psi_0}$  goes through a circuit made up of unitary gates $V_l^q$ and, possibly, measurement and feedback processes described by (measurement-conditioned) unitary operators $M_l^q$ arranged into $D$ layers. We assume ideal, unit-efficiency, measurements so that the computer's state remains pure. The $V_l^q \nobreak \coloneqq \nobreak e^{-i \theta_l^q H_l^q}$'s are parametrized by Hermitian operators $H_l^q$ and phases $\theta_l^q$. The index $q = \{1,\ldots, \mathcal{N}_l\}$ identifies the independent qubit (or set of qubits) on which each gate and measurement acts, and the index $l = \{1, \ldots, D\}$ identifies the circuit's layer. 
The total number of gates is
%\begin{equation}\label{eq:ngates}
$N_G \coloneqq \sum_{l=1}^D \sum_{q = 1}^{\mathcal{N}_l} 1 = \sum_{l=1}^D\mathcal{N}_l$. %\end{equation}

\begin{figure}
  \centering  
\includegraphics[trim=00 00 00 00,width=0.5\textwidth]{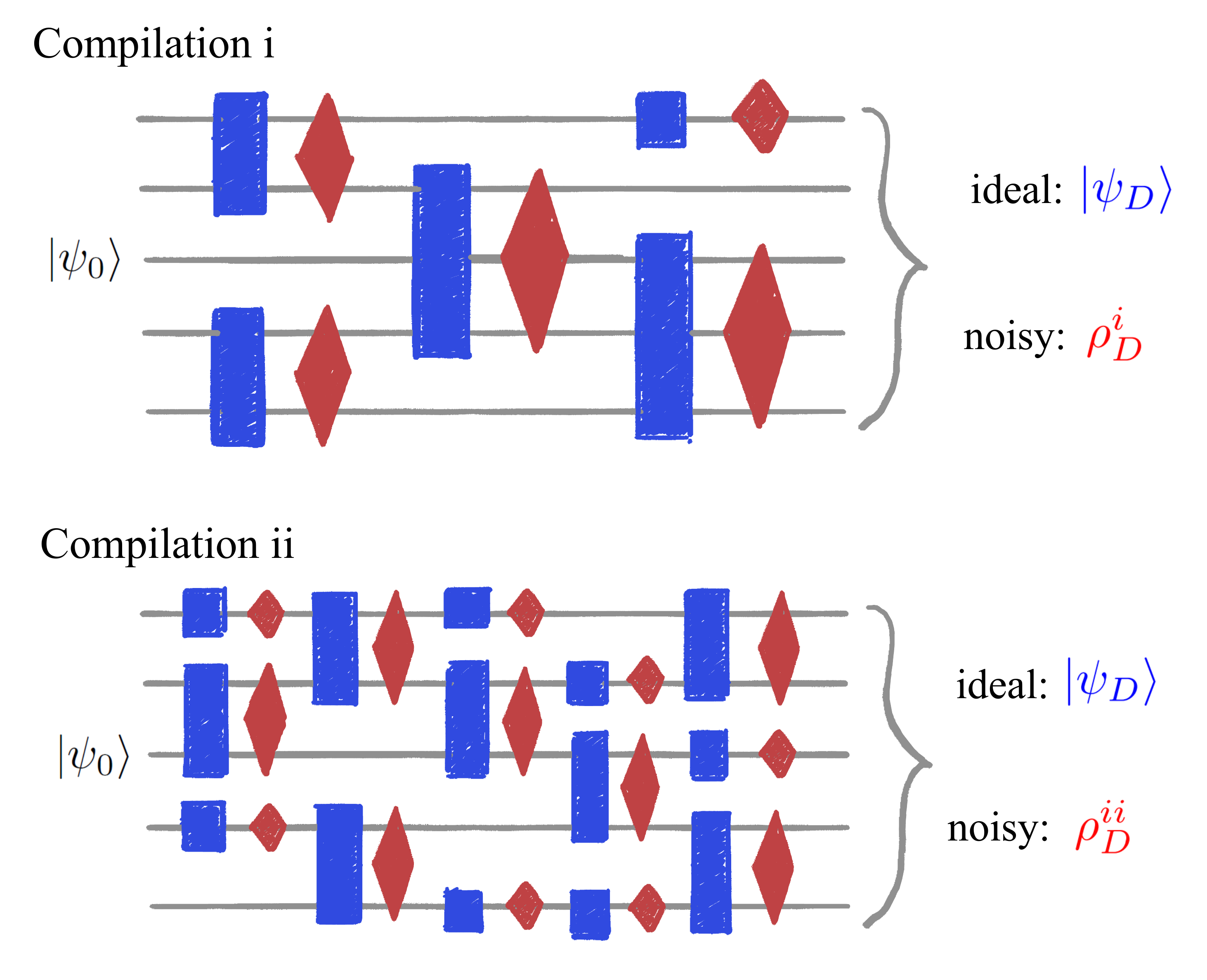}
%\\ \includegraphics[trim=00 00 00 00,width=0.47\textwidth]{fig11.pdf}
\caption{ 
\textbf{When is an algorithm resilient to noise?} 
% The blue circuit represents an ideal implementation of a quantum algorithm defined by gates $V_l^q$ and conditioned unitaries $M_l^q$ that describe possible measurement and feedback processes. The ideal circuit involves $N_G$ gates to prepare a state $\ket{\psi_D}$.   Unfortunately, any implementation of an algorithm suffers from errors,  illustrated by solid red spurious gates. The noisy implementation yields a perturbed state $\ket{\delta \psi_D}$.  In this work, we characterize implementations of an algorithm that are resilient against the noise affecting the system. Since the noise depends on the physical device that runs an algorithm, our framework can identify algorithms better suited to a given quantum computing platform.
  The circuits represent two compilations of a quantum algorithm that, under ideal conditions, drive the computer to a state $\ket{\psi_D}$. The blue boxes represent ideal circuit gates, which can include measurement and feedback processes. 
  Unfortunately, any realistic implementation of an algorithm suffers from errors,  illustrated by red rhombi. The noisy implementations yield perturbed (mixed) states $\rho_D^i$ and $\rho_D^{ii}$. 
  In this work, we characterize the noise-resilience of an algorithm by quantifying how the noisy outputs ($\rho_D^i$ and $\rho_D^{ii}$) deviate from the ideal one ($\ket{\psi_D}$). The noise resilience depends on the compilation. Sometimes (perhaps counter-intuitively) compilations with more gates are more resilient than short ones. 
 Since the noise depends on the physical device that runs an algorithm, our framework could identify algorithms better suited to a given quantum computing platform. 
\label{fig:fig1}
}
\end{figure}

 Measurements of $\ket{\psi_D}$ return the computation's result. However, a realistic quantum computer will not prepare $\ket{\psi_D}$ exactly, since it must rely on imperfect gates and suffer from extraneous noise sources. Then, instead of Eq.~\eqref{eq:idealdynamics}, a noisy implementation yields the perturbed dynamics
\begin{align}
\label{eq:noisydynamics}
 \ket{\psi_0}\!\bra{\psi_0} \xrightarrow[\textcolor{red}{\textnormal{(noisy)}}]{\textnormal{layer $1$}} \rho_1 \cdots \xrightarrow[\textcolor{red}{\textnormal{(noisy)}}]{\textnormal{layer $l$}} \rho_l \cdots \xrightarrow[\textcolor{red}{\textnormal{(noisy)}}]{\textnormal{layer $D$}} 
    \rho_D,
\end{align}
where $\rho_l$ is the (generally mixed) state obtained after $l$ noisy circuit layers. 
This article's main aim is to characterize how noise affects a computation, as illustrated in Fig.~\ref{fig:fig1}.
%Characterizing how noise affects the computation's result is the main aim of this article.

We start by considering coherent errors and show how to analyze incoherent errors in the next section. We model coherent errors by unitary gates $e^{-i Q_l^q  \delta\theta_l^q}$. The $Q_l^q$s and $\delta\theta_l^q$s are Hermitian operators and phases that characterize the noise process acting on qubit sets identified by $q$ after the $l$-th layer of the algorithm. Such errors perturb the state $\ket{\psi_l}$ at layer $l$ to the (pure) perturbed state $\rho_l \nobreak\coloneqq\nobreak \ket{\delta \psi_l}\!\bra{\delta \psi_l}$, with $\ket{\delta \psi_l} \nobreak \coloneqq \nobreak \prod_{l'=1}^l \bigotimes_{q=1}^{\mathcal{N}_{l'}} e^{-i \delta\theta_{l'}^q Q_{l'}^q } M_{l'}^q V_{l'}^q \ket{\psi_0}$.
 %We illustrate the perturbed dynamics in Fig.~\ref{fig:fig1}. 
 As we detail below, this framework can model imperfect implementations of a gate, decoherence due to external influences of an environment, and depolarizing noise. Since the $Q_l^q$s and $H_l^q$s can act on different qubit sets, this framework can also describe crosstalk and idling errors~\cite{sarovar2020detecting}. 
Note that the framework described above allows for different number of error and circuit gates. For instance, one can set certain $\delta\theta_l^q$ or $\theta_l^q$ to zero to specify error-less gates or that more than one error follows a gate, respectively.

To quantify the resilience of an algorithm against the noises $Q_l^q$, 
we consider the following measure of the \emph{fragility} of an algorithm:
\begin{align}
\label{eq:robustnessA}
%\label{eq:robustness}
\robust_Q &\coloneqq \nobreak  2 \big(1\nobreak -\nobreak \big|\langle \psi_D\! \ket{\delta\psi_D}\!\big| \big).
\end{align}
Equation~\eqref{eq:robustnessA} is the squared Bures distance $\mathcal{D}^2_B$ between the ideal and perturbed final states ($\robust_Q$ is itself a distance~\cite{PhysRevA.71.062310} and is a function of the fidelity). When the Bures distance between two states is sufficiently small no measurement can effectively distinguish them~\cite{bengtsson2017geometry}.
Similar metrics to analyze the resilience of metrological and control protocols have been considered in Refs.~\cite{modi2016fragile, PhysRevA.107.032606, Argentino}.

We say that the implementation of an algorithm is resilient (and not fragile) to the noises $ Q_l^q$ if $\robust_Q = \epsilon$ for $\epsilon \ll 1$. The fragility is a relevant measure for algorithms that encode the result of a computation in the final state~\cite{watrous2018theory}. While we focus on the state-fragility throughout the main text, we also consider the fragility of averages $\langle C \rangle = \bra{\psi_D} C \ket{\psi_D}$ for a cost function $C$ in Appendix~\ref{app:frag-expectation}.

An algorithm's fragility against the perturbative noise processes $Q_l^q$ satisfies
\begin{align}
\label{eq:ResultRobustness}
\robust_Q &\approx \sum_{j,k = 1}^D \sum_{q = 1}^{\mathcal{N}_j}\sum_{r = 1}^{\mathcal{N}_k}  \cov_{\ket{\psi_0}} \Big( Q_j^q(t_j), Q_k^r(t_k) \Big) \, \delta\theta_j^q \delta\theta_k^r, 
\end{align}
to leading orders in $\delta \theta_j^q$. 
Here, $\cov_{\ket{\psi}}(A,B) \coloneqq \tfrac{1}{2}\bra{\psi} \{A,B\} \ket{\psi} -  \bra{\psi} A \ket{\psi}\!\bra{\psi} B \ket{\psi}$ is the covariance between two operators $A$ and $B$ evaluated in state $\ket{\psi}$. We denote operators evolving in the Heisenberg picture by $Q_j^q(t_j) \nobreak \coloneqq \nobreak \left( \prod_{l=1}^j \bigotimes_{r=1}^{\mathcal{N}_l} M_l^rV_l^r \right)^\dag Q_j^q \left( \prod_{l=1}^j \bigotimes_{r=1}^{\mathcal{N}_l} M_l^rV_l^r \right)$, where $t_j$ identifies the time elapsed between the start of the protocol and the $j$-th layer of the circuit. Note that these time-evolved operators are evolved with the unperturbed circuit.
The approximation error $R$ in Eq.~\eqref{eq:ResultRobustness} satisfies $
    |R| \leq \frac{M}{6} \big( \sum_{l=1}^{D} \sum_{q=1}^{\mathcal{N}_l}|\delta \theta_l^q|  \big)^3$, 
where $M \coloneqq \max \big| \frac{\partial^{3}}{\partial \vect{\delta \theta}}  ( |\!\bra{\psi_D} \delta \psi_D \rangle| ) \big|$ and $\frac{\partial^{3}}{\partial \vect{\delta \theta}}$ denotes any third-order partial derivative with respect to the $\delta \theta$s.
To prove Eq.~\eqref{eq:ResultRobustness}, we leverage techniques from quantum information geometry that allow characterizing how the parametrized state $\ket{\delta \psi_D} \nobreak \equiv \nobreak \ket{\delta \psi_D}\big(\{\delta \theta_l^q\}\big)$ changes as the $\delta \theta_l^q$s change~\cite{provost1980riemannian,liu2019quantum} (see Appendix~\ref{app-main}).

Equation~\eqref{eq:ResultRobustness} shows that an algorithm's noise resilience depends on its compilation (the sequence of gates $V_l^q$ and $M_l^q$) and the noise affecting it (the noise operators $Q_l^q$). Other rather general analyses of the effect of noise on quantum algorithms have been performed in Refs.~\cite{stilck2021limitations,stabilitysimulation2022,berberich2023robustness}. Reference~\cite{stabilitysimulation2022} upper bounds the effect of noise on analog quantum simulators.   References~\cite{stilck2021limitations} and~\cite{berberich2023robustness} derive upper bounds on the effect of noise on quantum algorithms. The former considers noise channels with a fixed point and shows that noise above a certain threshold forbids quantum algorithms from demonstrating computational advantage. The latter reference upper bounds the effect of coherent over- and under-rotation errors (in our notation, they consider the particular case $Q_l^q = H_l^q$). 

Unlike Refs.~\cite{stilck2021limitations,berberich2023robustness}, we consider arbitrary coherent errors (and, we will prove later, decoherent errors) in a perturbative regime. The bounds in Refs.~\cite{stilck2021limitations,stabilitysimulation2022,berberich2023robustness} capture worst-case scenarios, so they often yield overly pessimistic estimates on the effect of noise. Instead,  Eq.~\eqref{eq:ResultRobustness}  approximates the influence of perturbative noise. This means that our results can give a more accurate depiction of the noise magnitudes that a computation can withstand without significant effects on the outcomes, which we illustrate below. 
Furthermore, estimating (instead of bounding) algorithms' noise resilience allows identifying compilations that are better suited to survive a noisy implementation. 
We discuss this further after studying resilience against characteristic noise sources that often affect physical devices.

\section{Resilience against uncorrelated noises and decoherence}
\label{sec:robustness-incoherent}

\subsection{Resilience against uncorrelated noise}
Consider a noise process in which the error affecting the set of qubits $(l,q)$ is independent of the errors influencing another set $(k,r)$. We model this by assuming that the $\delta\theta_l^q$s are uncorrelated zero-mean random phases, i.e., they satisfy $\overline{\delta\theta_l^q \delta\theta_k^r} = \delta_{lk} \delta_{q r} \sigma_{lq}^2$,  where $\sigma_{lq}$ is the standard deviation of $\delta\theta_l^q$. We use $\overline{f}$ to denote averages of a function $f$ over noise realizations.

We characterize the fragility of an algorithm's implementation against uncorrelated noises
by averaging Eq.~\eqref{eq:ResultRobustness} over the noises $\delta\theta_j^q$:
\begin{align}
\label{eq:ResultMeanRobustnessA}
\overline{\robust_Q} &\approx \sum_{l= 1}^D \sum_{q = 1}^{\mathcal{N}_l}  \sigma_{lq}^2 \,\var_{\ket{\psi_{l}}} \big(Q_l^q\big) .
\end{align}
To derive Eq.~\eqref{eq:ResultMeanRobustnessA}, we use that an operator's variance is $\var_{\ket{\psi_{l}}} (Q_l) \nobreak = \nobreak\var_{\ket{\psi_0}} (Q_l(t_l))\nobreak=\nobreak\cov_{\ket{\psi_0}}\big(Q_l(t_l),Q_l(t_l)\big)$.

Eq.~\eqref{eq:ResultMeanRobustnessA} %and~\eqref{eq:ResultMenRobustnessC} 
illustrates how some compilations of a quantum algorithm can be more resilient against a particular noise source than others. Protocols that drive the computer's state through a path with small uncertainties in $Q_l^q$ are more resilient than implementations passing through states with large variances $\var_{\ket{\psi_{l}}} \big(Q_l^q\big)$ of the noise operators.
For example, as we demonstrate in the next section, adiabatic algorithms that force the computer's state to remain in a low-energy subspace (and, therefore, to small energy uncertainties), are particularly resilient to over-rotation errors and energy decoherence.

Eq.~\eqref{eq:ResultMeanRobustnessA} also informs a characterization of the noise regimes within which an algorithm is resilient. 
For example, consider an algorithm where single-qubit Pauli noise operators act on all qubits at each layer with uniform noise intensity $\sigma_{lq} \eqqcolon \sigma$. Typically, non-idle qubits involved in a computation are entangled with neighboring qubits~\cite{jozsa2003role}. This leads to variances $\var_{\ket{\psi_{l}}} \big(Q_l^q\big) \sim 1$. The same approximate variances hold for random single-qubit Pauli noise operators, regardless of the qubits' entanglement.
Then, Eq.~\eqref{eq:ResultMeanRobustnessA} leads to $\overline{\robust_Q} \sim D n \sigma^2$, where $n$ is the number of qubits. In such a scenario, the noise-per-gate needs to scale as $\sigma^2 \sim \epsilon/(Dn)$ for a small $\epsilon$ if the algorithm is to be noise-resilient. 

For comparison, for over-rotation or under-rotation errors ($Q_l^q \equiv H_l^q$), Theorem II.2 of Ref.~\cite{berberich2023robustness} implies that $\robust_H \geq (\max_{l,q} \delta \theta_l^q)^2 \big( \sum_l \sum_q \|H_l^q\| \big)^2 \sim \sigma^2 n^2 D^2$ in the scenario described above. While their result holds beyond the perturbative regime, it yields an overly conservative estimate of how noise must scale with the circuit size for an algorithm to be robust: $\sigma^2 \sim \epsilon/(D n)^2$.

Thus, for small errors, the approximate figure of merit for an algorithm's resilience in Eq.~\eqref{eq:ResultMeanRobustnessA} can yield widely different results than the worst-case bounds derived in the literature.
Finally, observe that the more general expression~\eqref{eq:ResultRobustness} allows, in principle, for higher error thresholds for correlated noise, since the cross terms $\cov \big( Q_j^q(t_j), Q_k^r(t_k) \big)$ can be negative.

\subsection{Resilience against incoherent noise} \label{subsec:incohe}
Most noise causes decoherence of a quantum computer's state, mixing it as it evolves, as depicted by Eq.~\eqref{eq:noisydynamics}. Dephasing and bitflip noises are examples of incoherent errors that often affect qubits. Leveraging the fact that we are working in the perturbative regime, the formalism introduced above can also describe incoherent errors, as we describe next.

Under dephasing noise a qubit's state $\rho_{\textnormal{qubit}}$ evolves to $\rho_{\textnormal{qubit}}' \nobreak= \nobreak \left( 1 \nobreak- \nobreak\frac{p}{2} \right)\rho_{\textnormal{qubit}}  \nobreak+ \nobreak \frac{p}{2} \left( Z \rho_{\textnormal{qubit}} Z \right)$, where $Z$ is a Pauli matrix and $p$ is the probability of an error occurring~\cite{preskill2015lecture}. Averaging over uncorrelated coherent noise $e^{-i \delta \theta Q }$ described by $Q = Z$ and $\sigma^2 =  \overline{\delta \theta^2}  \eqqcolon  p/2$ models the effect of dephasing noise. Similarly, bitflip noise,  depolarizing noise, and decoherence that affect sets of qubits can be described by averaging over coherent noise with suitably chosen $Q$s. These statements are proven in Appendix~\ref{app-incoherent}.

In the general case, upon being affected by a sequence of decoherent noisy layers throughout the computation, the system will end in the mixed state $\rho_D = \overline{\ket{\delta \psi_D}  \!\bra{\delta \psi_D}}$ instead of $\ket{\psi_D}\!\bra{\psi_D}$. The overlap of these states is 
\begin{align}
\label{eq:incoherentoverlap}
    \tr{\rho_D \ket{\psi_D}\!\bra{\psi_D}} &= 1 - \overline{\robust_Q}  + \overline{\robust_Q^2}/4 \approx 1 - \overline{\robust_Q} \nonumber \\ &\approx 1 - \sum_{l= 1}^D \sum_{q = 1}^{\mathcal{N}_l}  \sigma_{lq}^2 \,\var_{\ket{\psi_{l}}} \big(Q_l^q\big),
\end{align}
to leading order in the perturbation parameters, which we prove by using Eqs.~\eqref{eq:robustnessA},~\eqref{eq:ResultRobustness}, and~\eqref{eq:ResultMeanRobustnessA}.

\begin{figure*} 
  \centering  
  \begin{tabular}{c}
    \begin{subfigure}[b]{.26\textwidth}
        \captionsetup{skip=30pt, slc=off, margin={-15pt, 0pt}}
        \caption{Planar Surface $d$ = 2}
        \includegraphics[trim=15cm 10cm 15cm 5cm, width=\textwidth]{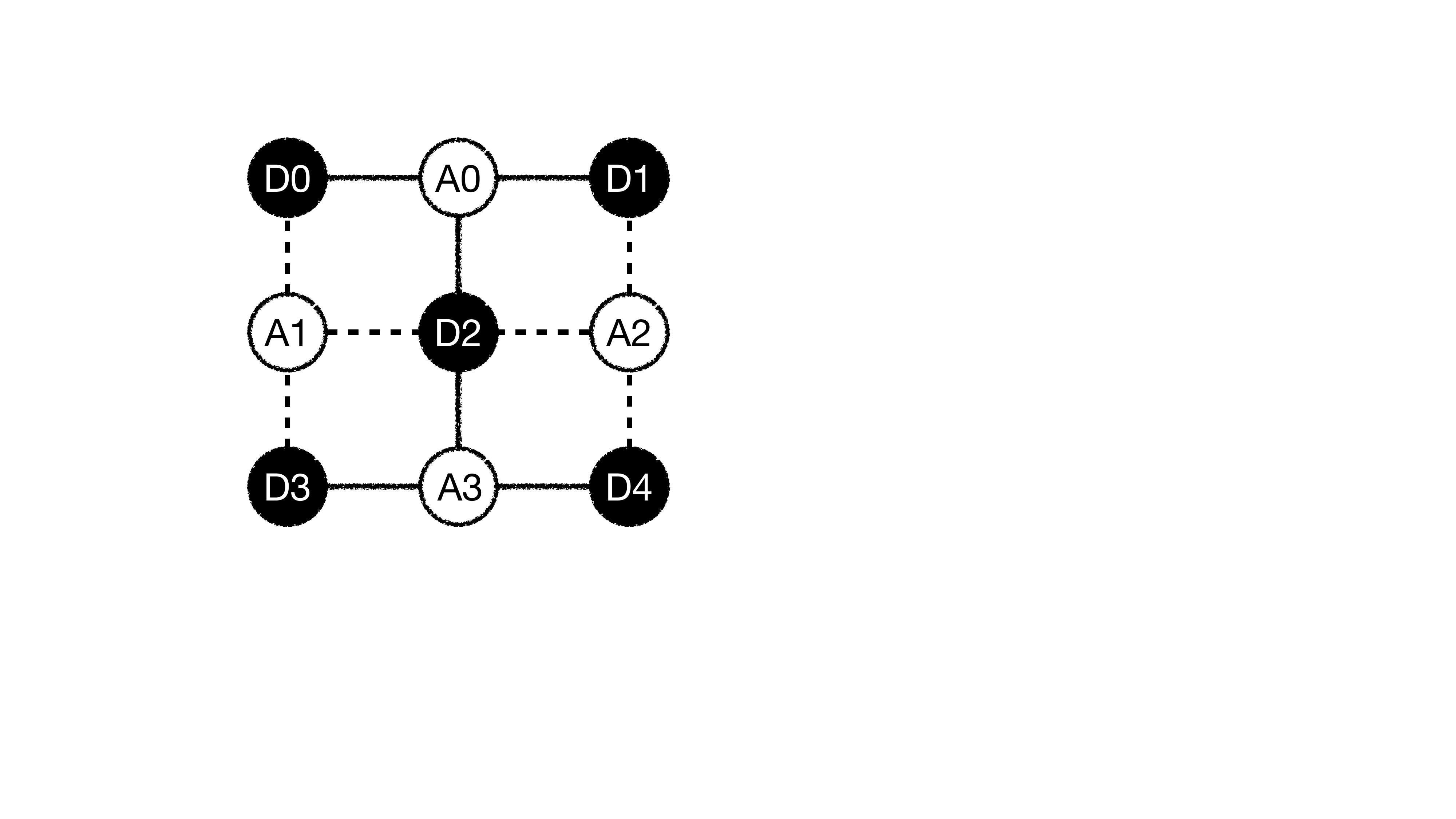}
    \end{subfigure} \\
    \begin{subfigure}[b]{.26\textwidth}
        \captionsetup{skip=-10pt, slc=off, margin={-15pt, 0pt}}
        \caption{$XZZX$ $d$ = 2}
        \includegraphics[trim=15cm 4cm 15cm 5cm, width=\textwidth]{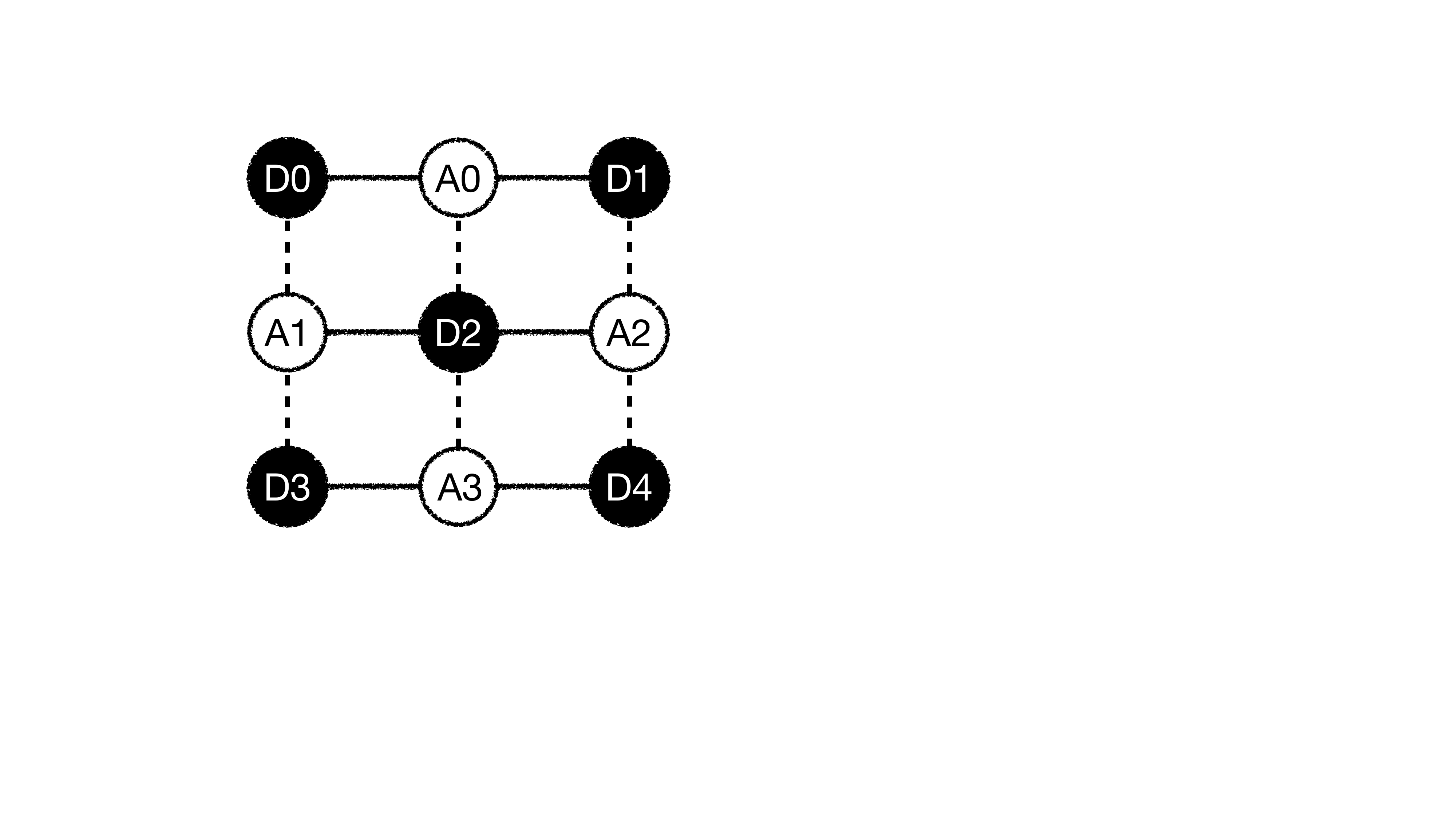}   
    \end{subfigure} 
  \end{tabular} 
  \hspace{0.5cm}
    \begin{subfigure}[t]{.5\textwidth}
    \includegraphics[trim=5cm 7.25cm 00 0cm, width=\textwidth]{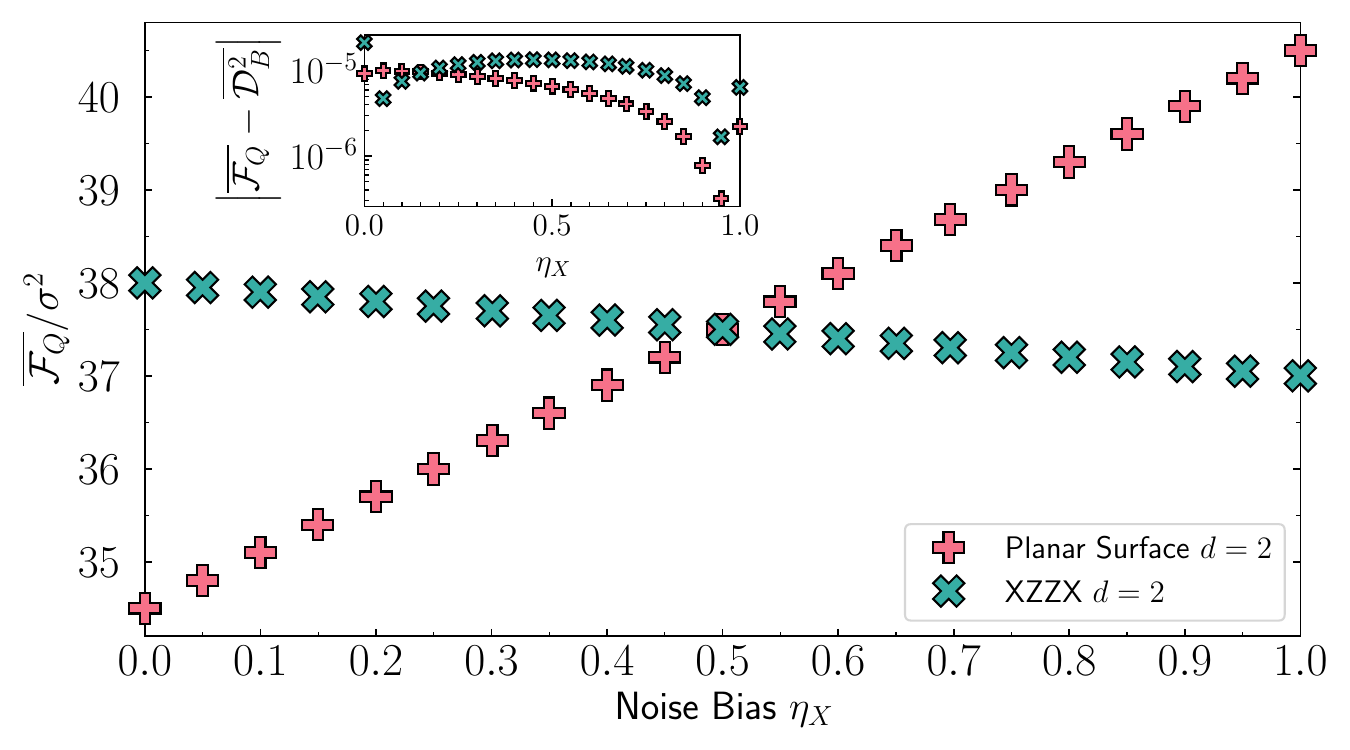} 
    \captionsetup{width=\textwidth, skip=-91pt, margin={-345pt,0pt}}
    \caption{}  
    \end{subfigure}
\caption{ 
\textbf{Identifying resilient compilations of an algorithm.} 
In an ideal scenario, the $2$-dimensional planar and $XZZX$ surface codes can detect errors in a logical qubit due to a single error affecting a physical qubit. 
But, what if the error detection circuitry is affected by errors? Can different detection circuits operate more 
reliably under different types of noise?
We evaluate the normalized noise-averaged fragility $\overline{\robust_Q}/\sigma^2$ of parity-check circuits for the planar and $XZZX$ surface codes, at distance 2. 
The code circuits are diagrammatically presented in (a) and (b). Solid and dotted lines represent controlled $X$ and $Z$ operations, controlled on ancillary qubits (A) and targeting data qubits (D).
We use a single-qubit Pauli noise model with a uniform error rate $p = 2\sigma^2$. The noise model contains a bias parameter, $\eta_X$,
which interpolates between pure dephasing ($\eta_X = 0$), bitflip ($\eta_X = 1$) and equal probability bit phase/flip noises ($\eta_X = 0\text{.}5$). 
As circuit input, we consider an arbitrary state $\ket{\psi_0} = \alpha \ket{\widetilde{0}} + \beta \ket{\widetilde{1}}$ of the 
logical qubit, where $\ket{\widetilde{0}}$ and $\ket{\widetilde{1}}$ span the logical codespace. The results are independent of $\alpha$ and $\beta$. 
We find that the noise resilience is code-dependent: the planar surface code circuit is more resilient to pure dephasing noise while being worse at coping with pure bitflip noise, with the converse being true for the XZZX code circuit. The resilience of both detection circuits coincides for equally-weighted bit/phase flip noise, $\eta_x = 1/2$. 
The inset of (c) quantifies the approximation error between the expression for fragility presented in Eq.~(\ref{eq:ResultMeanRobustnessA}) (which requires no averaging) and the average of Eq.~(\ref{eq:robustnessA}) (labeled as $\overline{\mathcal{D}^2_B}$, which requires exact statevector simulation to evaluate) over multiple noise instances with single-qubit error rate $p=0\text{.}0001$.
 We find that the 9.21\% normalized fragility gap at $\eta_X = 0$ corresponds to average fidelity gaps of 0.0320$\pm$0.005\%, 0.329$\pm$0.05\% and 3.39$\pm$0.4\% for total single qubit error rates of $p = 0\text{.}0001, 0\text{.}001, 0\text{.}01$ respectively. 
 %This example illustrates the potential of Eq.\eqref{} to asses the suitability 
\label{fig:fig2}
}
\end{figure*}

Eq.~\eqref{eq:incoherentoverlap} provides a straightforward way to identify compilations of an algorithm that are better suited to withstand certain incoherent noises. 
For example, consider two quantum error detection parity-check circuits, the planar surface code~\cite{RoffeQEC2019} and the $XZZX$ surface code~\cite{bonilla2021xzzx} at distance $d=2$. 
These circuits are presented diagrammatically in Fig.~\ref{fig:fig2}(a) and~\ref{fig:fig2}(b). 
In an ideal scenario, both codes rely on unitary gates and measurements to detect a single error (up to a threshold error rate) affecting a data qubit within a logical qubit.
In Fig.~\ref{fig:fig2}(c), we study the codes' resilience to implementation errors that uniformly affect all physical qubits. 
We use a single-qubit Pauli noise model that can interpolate between phase and bit-flip noises (see details in Appendix~\ref{app-biased}).
We find that the planar surface code circuit is less fragile to noise dominated by phase flip errors than the $XZZX$ code circuit but more fragile to noise dominated by bit-flip errors.

In the previous example, we studied the noise fragility of circuits under incoherent noise by averaging over multiple instances of coherent noise. However, we highlight that Eq.~\eqref{eq:ResultMeanRobustnessA} enables evaluation of the resilience of stabilizer code circuits to arbitrary small coherent errors using the stabilizer formalism. This sidesteps potentially expensive noisy statevector simulations and relaxes assumptions on the form of coherent errors made by direct simulations of error correction protocols~\cite{bravyi2018correcting}. Instead, the noise-resilience of the code circuitry depends on the variances of the noise operators in the evolving unperturbed state $\ket{\psi_l}$ of the qubits.
Equations~\eqref{eq:ResultMeanRobustnessA} and~\eqref{eq:incoherentoverlap} thus provide simple figures of merit to discriminate the noise-resilience of different digital circuits.

\subsection{Noise-resilience of analog quantum algorithms}
Analog quantum algorithms replace the sequence of gates $V_{l}^q$ by an explicitly time-dependent Hamiltonian $H_t$. This describes computing by quantum annealing and protocols for quantum simulation~\cite{RevModPhys.80.1061}. 
By taking a time-continuum limit in Eqs.~\eqref{eq:ResultRobustness} and~\eqref{eq:ResultMeanRobustnessA}, we can characterize the noise-resilience of analog quantum algorithms. 

We model the noise affecting an analog algorithm by a (possibly time-dependent) Hermitian noise operator $Q$ and a stochastic noise function $\xi_t$~\cite{jacobs2010stochastic}. The analog computer's state then evolves under the noisy Hamiltonian $H_t \nobreak+\nobreak \xi_t Q$. We assume that $\gamma_t$ characterizes the noise's intensity, and $\overline{\xi_t \xi_{t'}} \nobreak=\nobreak \gamma_t \, \delta(t-t') $. Upon averaging over noise realizations, such noisy dynamics is equivalent to a Lindblad master equation $\dot \rho_t \nobreak=\nobreak -i[H_t,\rho_t] \nobreak-\nobreak \gamma_t [Q,[Q,\rho_t]]$ governing dynamics of the analog computer's state $\rho_t$~\cite{ChenuPRL2017}.

The average fragility of a $T$-runtime analog algorithm affected by continuous uncorrelated noise satisfies
\begin{align}
\label{eq:meanResultRobustnessCont}
\overline{\robust_Q} &\approx \int_0^T \gamma_t \,  \var_{\ket{\psi_t}} \big( Q \big) \,  dt.
\end{align}
Let us illustrate the use of Eq.~\eqref{eq:meanResultRobustnessCont} to identify noise-resilient compilations of a simple state-preparation protocol. Consider flipping two qubits from state $\ket{00}$ to $\ket{11}$. Two unitaries that accomplish this are $V_a = e^{-i \pi/2 H_a}$ and $V_b = e^{-i \pi H_b}$, with $H_{a} \nobreak=\nobreak -i\Big(\ketbra{00}{11}-\ketbra{11}{00}\Big)$ and $H_b \nobreak=\nobreak \frac{1}{2} \left( Y \otimes \id + \id \otimes Y \right)$. $X$, $Y$, and $Z$ denote Pauli matrices. Consider noise processes (i) and (ii) described by operators $Q_i \coloneqq X \otimes X$ and $Q_{ii} \coloneqq \frac{1}{4}(\id - Z) \otimes (\id - X)$, respectively. 
We show in Appendix~\ref{app-simpleexample} that 
$\overline{\robust_{Q_i}^a} < \overline{\robust_{Q_i}^b}$ but $\overline{\robust_{Q_{ii}}^a} > \overline{\robust_{Q_{ii}}^b}$. That is, the implementation by $H_{a}$ is more resilient to noise (i) while the implementation by $H_{b}$ is better suited to avoid errors due to the noise (ii). 
Equation~\eqref{eq:meanResultRobustnessCont} and its digital version~\eqref{eq:ResultMeanRobustnessA} thus provide simple ways to evaluate compilations' noise-resilience which do not require simulating the noise.

One may wonder how an algorithm's resilience relates to its runtime. Intuitively, the longer the runtime the more errors are accumulated. However, we can use Eq.~\eqref{eq:meanResultRobustnessCont} to show this intuition wrong. Consider, for simplicity, a system driven by a fixed Hamiltonian $H$ and affected by an error operator $Q = H$ at constant noise intensity, $\gamma_t \equiv \gamma$. This models energy decoherence, for instance, due to the use of an imperfect clock to determine the duration of a computing schedule~\cite{EgusquizaClocks1999, GambiniClocks2004, HuberClocks2023}. Rescaling the Hamiltonian $H' = a H$ by a factor $a>1$  diminishes an algorithm's runtime to $T' = T/a$. Great, it computes faster! Unfortunately, the rescaling also increases the algorithm's fragility, $\robust_{\ket{\psi}}^{H'} = \gamma \int_0^{T/a} \var_{\ket{\psi_t}} (aH) dt = a\robust_{\ket{\psi}}^{H}$. In this simple example, decreasing the computation time leads to a less resilient algorithm. We will show that this result can be formalized as a resilience-runtime tradeoff relation that constrains any algorithm under more general noise processes.

\section{Resilience--runtime tradeoff relations}
\label{sec:tradeoffs}
While intuition suggests that the least number of gates in a circuit the better, the discussion above shows that this expectation can be misleading. Sometimes, compilations of an algorithm that involve more gates (which often mean longer runtime) are more resilient than compact ones. 

The tradeoff relations 
\begin{subequations}
\label{eq:tradeoffState}
\begin{align}
\label{eq:tradeoffStateCircuit}
    N_G \, \overline{\robust_Q} &\geq \min_{lq} \left( \frac{\sigma_{lq}}{\theta_{l}^{q}} \right)^2 \length_Q^2, \qquad \textnormal{and}\\
    \label{eq:tradeoffStateCont}
    T \overline{\robust_Q} &\geq \min_t \gamma_t \,\, \mathcal{L}_Q^2,
\end{align}
\end{subequations}
constrain the average fragility and the number of gates $N_G$ or runtime $T$ of digital and analog algorithms, respectively~\footnote{As described in the paragraph that follows Eq.~\eqref{eq:noisydynamics}, the framework used in this article allows for different number of circuit gates and error gates. A similar tradeoff relation to Eq.~\eqref{eq:tradeoffStateCircuit} holds in this case, with $N_G$ being the maximum between the circuit and error gates.}.
The lower bounds in the tradeoff relations involve   
\begin{align}
\label{eq:distance}
   \length_Q \coloneqq \sum_{l q} |\theta_{l}^{q}| \sqrt{\var_{\ket{\psi_l}} (Q_l^q)},
\end{align}
for digital circuits, and $\mathcal{L}_Q \nobreak\coloneqq\nobreak \int_0^T \!\! \sqrt{\var_{\ket{\psi_t}} (Q)} dt$ for time-continuous algorithms (see proof in Appendix~\ref{app:tradeoff}).

\begin{figure}
  \centering  
  \includegraphics[trim=00 00 00 00,width=0.47\textwidth]{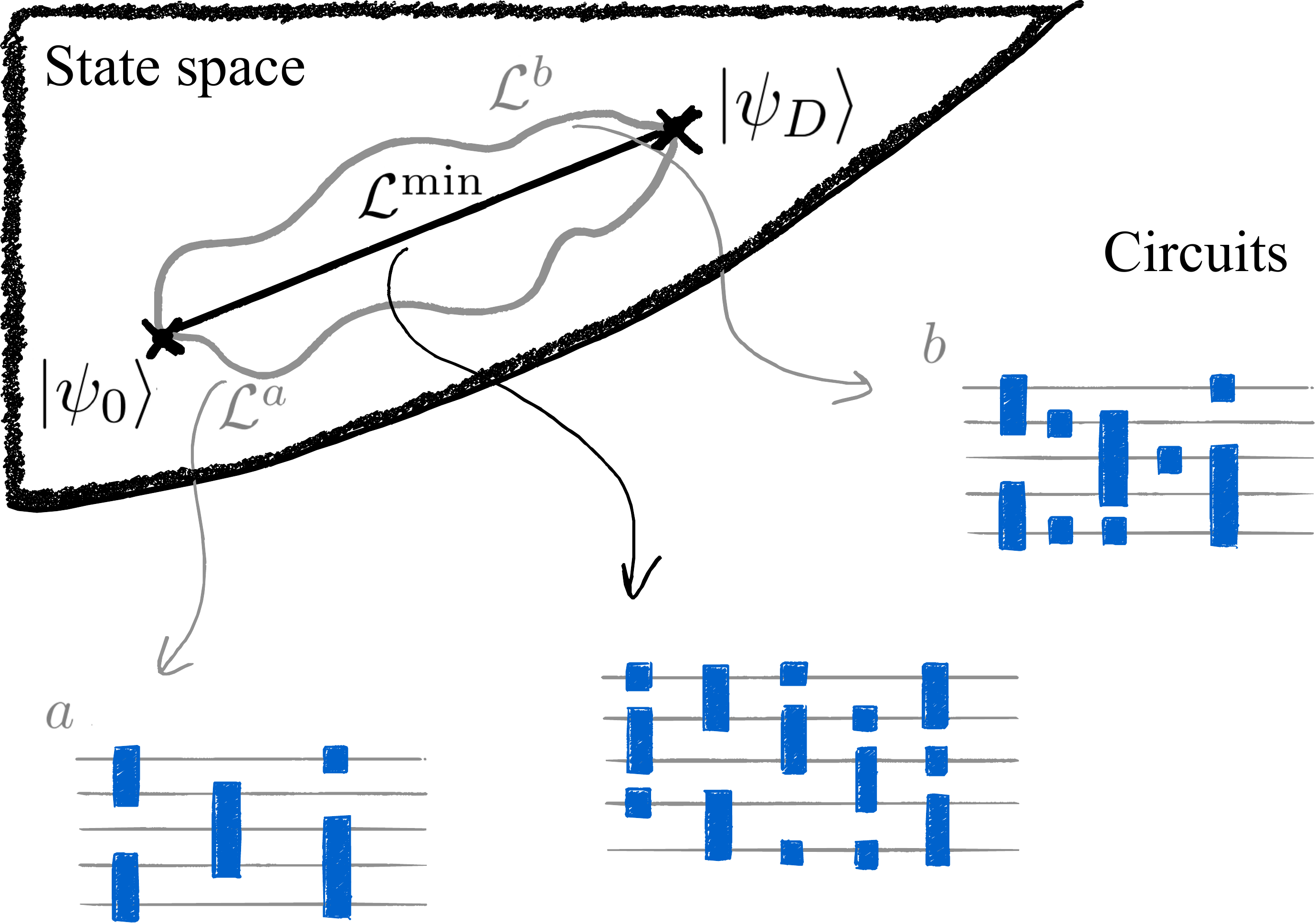} 
\caption{ 
\textbf{Resilience--runtime tradeoff relations.} 
A circuit's fragility and number of gates (or runtime in the case of analog algorithms) are constrained by Eq.~\eqref{eq:tradeoffState}. When a circuit's gates are affected by over/under-rotation errors (e.g., due to timing inaccuracies), the lower bound in the tradeoff relations depends on the path length $\mathcal{L}$ traversed by the computer's state. Different circuits that implement the same computation will generally lead to different path lengths. 
The tradeoff relations suggest that shorter paths through state space may allow for smaller $N_G \robust_H$ than longer paths. Moreover, for a given length $\mathcal{L}$, Eq.~\eqref{eq:tradeoffState} constrains the minimum number of gates that must be applied to attain a desired noise-resilience.
\label{fig:fig3}
}
\end{figure}

Eq.~\eqref{eq:tradeoffState} says that, for a fixed $\length_Q$, the number of gates in a circuit or an algorithm's runtime cannot be too small if one aims for a resilient algorithm. It also suggests that, in some cases, computing for longer may help avoid errors due to noisy dynamics. We confirm this to be the case below.
References~\cite{AccuracyAnnealingLidarPRA2010,nakajima2024speed} suggest other potential benefits of longer computation times, in their case, in terms of improved computational accuracy.

To interpret the lower bound in Eq.~\eqref{eq:tradeoffState}, note that the length of the path traveled by a pure state driven by a Hamiltonian $H_t$ is $\mathcal{L} \nobreak \coloneqq \nobreak \int_0^T \!\sqrt{\var_{\ket{\psi_t}} ( H_t  )} dt $~\cite{bengtsson2017geometry, GeometricQSLPRX2016}. For a time-independent Hamiltonian with a duration $T \nobreak=\nobreak\theta $, $\mathcal{L} \nobreak=\nobreak \theta\sqrt{\var_{\ket{\psi_0}}(H)}$.
Then, each summand in $\length_Q$ has a geometric interpretation as the length of a hypothetical path in Hilbert space if the state were driven by noise operators $Q_l^q$ for durations $|\theta_{l}^{q}|$. 

The geometric quantity $\mathcal{L}_Q$ has an elegant meaning when over/under-rotation errors or energy decoherence affect the system, i.e., for $Q_l^q = H_l^q$ or $Q = H_t$. Then, $\mathcal{L}_Q \equiv \mathcal{L}$ coincides with the path length covered by the system's state $\ket{\psi_l}$ throughout the ideal evolution~\cite{bengtsson2017geometry, GeometricQSLPRX2016}. An algorithm's compilation that drives the state through a short path is less constrained---the products $N_G \overline{\robust_H}$ or $T \overline{\robust_H}$ in Eq.~\eqref{eq:tradeoffState} can be smaller---than one with a long path. We illustrate the resilience--runtime tradeoff relation in Fig.~\ref{fig:fig3}.

Computing faster, for instance, by rescaling $H_t$ as described in the previous section, decreases an algorithm's runtime but can also result in worse noise resilience. The distance $\mathcal{L}$ is independent of the speed at which the system evolves, so Eq.~\eqref{eq:tradeoffState} implies that an arbitrary decrease in $T$ must come at the expense of worse noise-resilience against over/under-rotation errors or energy decoherence. Similarly, significant runtime enhancements by protocols that exploit shortcuts to adiabaticity~\cite{AdCShortcuts2010, AdCShortcuts2013, Polkovnikov2017} 
%, MugaShortcuts2019}
or fast-forwarding~\cite{FF2017, FF2021} to rapidly drive the state through a fixed $\mathcal{L}$-length path must result in worse noise-resilience.

\begin{figure}
  \centering  
  \includegraphics[trim=10 00 00 00,width=0.5\textwidth]{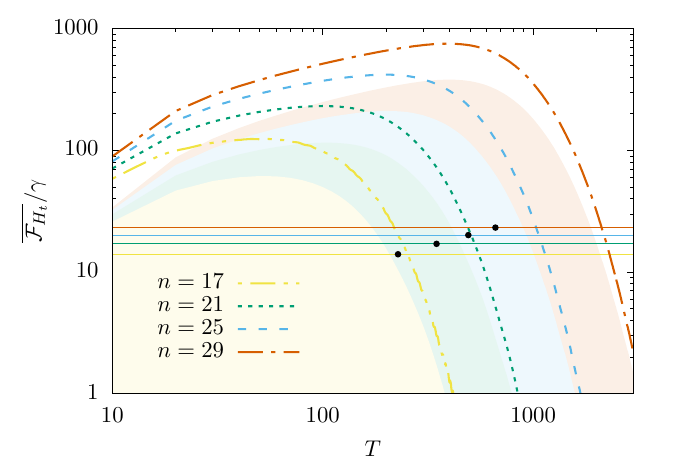} 
\caption{ 
\textbf{Resilience of adiabatic and optimized compilations for the $p$-spin model.} 
Plot of the fragility [Eq.~\eqref{eq:meanResultRobustnessCont} with $Q=H_t$ and constant $\gamma$] as a function of runtime for adiabatic compilations of the $p$-spin model with a linear annealing ramp.
The $p$-spin model consists of the time-dependent $n$-qubit Hamiltonian $H_t \nobreak=\nobreak (1 \nobreak-\nobreak g_t) H_0 \nobreak+\nobreak g_t H_1$, where  $H_0 = -  \sum_{j=1}^n X_j/2 $ and $H_1 = - \big( \sum_{j=1}^n Z_j \big)^p/(2 n^{p-1})$, and $p > 0$ is a free parameter. The schedule $g_t$ is chosen so that the state approaches the ground state of $H_1$ after a time $T$.
These simulations are for $p=3$ which is known to have exponential adiabatic runtime scaling \cite{pspin} but also has a polynomial time bang-bang schedule which exactly reaches the ground state (for odd $n$)~\cite{pspinQAOA}.
The dashed curves show the fragility of these adiabatic schedules. The solid horizontal lines of the same color represent the fragility of the analytically optimized bang-bang schedule with the black dots corresponding to the runtime those bang-bang schedules would take.
The shaded regions depict fragility values forbidden by the bound 
%on $\mathcal{L}_Q^2/T$ obtained from 
in Eq.~\eqref{eq:tradeoffStateCont} applied to the correspondingly-colored compilations.  From bottom to top, these shaded regions increase in $n$, corresponding to the procedures they end up tracking for large $T$.
Key to note here is that for the adiabatic schedules, there is always a runtime such that the fragility is below some threshold, meaning that increased runtime is all that is needed to reduce fragility, unlike the optimized schedules which run at fixed runtime and fragility.
\label{fig:fig4}
}
\end{figure}

In variational quantum algorithms, a popular approach to solving computational tasks, compilations are typically optimized to minimize circuit depth or computation time~\cite{farhi2015quantum, bauer2020quantum, cerezo2021variational}. This can significantly improve time performance. For example, for the $p$-spin model, an optimized algorithm can compute exponentially faster than an adiabatic annealing one~\cite{pspinQAOA}. Intuitively, given the exponentially longer times over which errors can accumulate in the adiabatic algorithm, one may expect it to be less resilient than the optimized one. 
 However, adiabatic algorithms drive the computer's state through paths close to the instantaneous ground state of a time-dependent Hamiltonian $H_t$. Thus, for noise described by $Q = H_t$ (e.g., due to timing inaccuracies in the dynamics), the variance $\var_{\ket{\psi_t}} (H)$ in the fragility~\eqref{eq:meanResultRobustnessCont} is small during adiabatic dynamics, improving noise-resilience. 
%We show in Fig.~\ref{fig:fig4} that this is not the case; the adiabatic algorithm can be more noise-resilient despite the exponentially longer runtime. 
We show that the adiabatic algorithm can be more noise-resilient than an optimized one despite the exponentially longer runtime in Fig.~\ref{fig:fig4}. 
Meanwhile, the optimized algorithm can be more resilient to other noises. % (see Appendix~\ref{app:p-spin}).  
This discussion underlines the need for tailoring an algorithm's compilation to the noise affecting the computational device.

\section{Discussion}

The main results in this article [Eqs.~\eqref{eq:ResultRobustness} through~\eqref{eq:tradeoffState}] can be leveraged to identify noise-resilient compilations of an algorithm. 
Naturally, in a thriving field with tens of thousands of publications, one can find many previous examples of work on algorithm compilation and noise-resilience. Reference~\cite{chong2017programming} reviews the literature on hardware-informed algorithm compilation.  As examples, Ref.~\cite{venturelli2018compiling} considers hardware-aware circuit compilation via temporal planners to minimize runtime, Ref.~\cite{venturelli2019quantum} introduces software tools to find short-runtime compilations, and Ref.~\cite{CalderonVargas2023quantumcircuit} introduces a method to identify circuit layers that make an algorithm sensitive to noise.  Compilations optimized to be noise-robust, resource-efficient, or precise have been studied in more detail in variational algorithms~\cite{SimonPRX2017VarSimul,
%compilationQAOA,
PhysRevA.104.022403, MurphyNiu2019, PhysRevA.105.052414, jones2022robust, SucharaOptimizingGates}, along with the characterization of noise-resilience of learning algorithms~\cite{weber2021optimal, Sharma_2020,
skolik2023robustness}. 

Our contribution is a general framework to characterize noise resilience that can be straightforwardly applied given knowledge of an ideal circuit and a model of the noise affecting the quantum computer. 
%We derived figures of merit to quantify the noise-resilience of quantum algorithms. 
To do so, we relied on tools from quantum information geometry (namely, the Fisher or Fubini-Study metric) that describe how a parametrized state changes under small parameter changes. 
(Similar tools are used in Refs.~\cite{modi2016fragile,Argentino,PhysRevLett.131.210601, PhysRevLett.130.160802} to analyze the resilience of metrological and quantum control protocols and to bound the minimum cost to perform quantum error mitigation, respectively.) We focused on noise's effect on the final state of the computer, but our results extend to the noise resilience with respect to arbitrary cost functions (Appendix~\ref{app:frag-expectation}).

The reader may wonder if a perturbative regime is powerful enough to study the resilience of quantum algorithms in realistic scenarios. The answer is likely to depend on the context.
We expect our approach to be more naturally applied in a modular way, for instance, to identify gate compilations tailored to resiliently run despite a certain noise affecting an experimental platform. Intuitively, choosing compilations that are resilient to small errors is more likely to lead to better overall results than noise-agnostic compilations. For a different approach see Refs.~\cite{stilck2021limitations,stabilitysimulation2022,berberich2023robustness}, which derive worst-case bounds that hold away from a perturbative regime. However, such bounds often give overly pessimistic estimates for the error rates necessary to yield resilient algorithms. In practice, an approximate measure may sometimes be preferred over a worst-case bound.

Moreover, focusing on a perturbative regime leads to resilience metrics [Eqs.~\eqref{eq:ResultRobustness} through~\eqref{eq:meanResultRobustnessCont}] that depend on the unperturbed state of the computer under ideal conditions. This allows evaluating resilience without simulating noisy dynamics, which we exploited in Figs.~\ref{fig:fig2} and~\ref{fig:fig4} to identify resilient compilations of a digital error detection protocol and of an analog state preparation algorithm. It would be interesting to use the resilience metrics to analyze other algorithms, in particular, to identify compilations that better suit the noises that affect competing computing platforms~\cite{bluvstein2024logical,AbsenceofBarrenPRX2021,
debroy2020logical,
IBMkim2023evidence,
IBMvan_den_Berg_2023}. In parallel, it would be interesting to characterize how the error thresholds change with the choice of noise-tailored error correction protocols in the presence of biased noise.
And, while our examples focus on uncorrelated errors, future work using Eq.~\eqref{eq:ResultRobustness} could shed light on the role of spatially and temporally correlated noises and how to tailor algorithm compilations to such noises.

We proved a tradeoff between an algorithm's noise resilience and its number of gates or runtime. This tradeoff relation depends on a geometric quantity that quantifies path lengths in Hilbert space. There are direct connections between Eq.~\eqref{eq:tradeoffState} and other geometric approaches to study quantum algorithms. Brown and Susskind's circuit complexity, $\mathcal{C} \coloneqq \min_{\textnormal{compilations}} N_G$, is defined by the minimum number of gates needed to implement a unitary~\cite{PhysRevD.97.086015, Haferkamp_2022}. Minimizing Eq.~\eqref{eq:tradeoffStateCircuit} over circuit realizations then implies a tradeoff relation between Brown and Susskind's circuit complexity and the noise-resilience of the minimal circuit in terms of a path length---a complexity-resilience tradeoff relation. It would be interesting to also explore connections to Nielsen's notion of circuit complexity~\cite{nielsen2005geometric, Nielsen, brown2023quantum}. Regardless, the tradeoff relations~\eqref{eq:tradeoffState} and the example in Fig.~\ref{fig:fig4} show that minimizing the number of gates or the runtime can lead to fragile algorithms; perhaps a more wholesome notion of complexity should incorporate a circuit's noise-resilience.

\vspace{6pt}
{\fontsize{11}{13}
\noindent\textbf{Acknowledgments}}\\
We thank Victor Albert for comments on the manuscript.  
This material is based upon work supported by the U.S. Department of Energy, Office of Advanced Scientific Computing Research, Accelerated Research for Quantum Computing program, Fundamental Algorithmic Research for Quantum Computing (FAR-QC) project. 
We acknowledge support by the Laboratory Directed Research and Development (LDRD) program of Los Alamos National Laboratory (LANL) under project number 20230049DR, and Beyond Moore’s Law project of the Advanced Simulation and Computing Program at LANL managed by Triad National Security, LLC, for the National Nuclear Security Administration of the U.S. DOE under contract 89233218CNA000001. 
%The work at LANL was carried out under the auspices of the US DOE and NNSA under contract No.~DEAC52-06NA25396. 
J.B.~was supported in part by the DoE ASCR Accelerated Research in Quantum Computing program (award No.~DE-SC0020312), DoE ASCR Quantum Testbed Pathfinder program (awards No.~DE-SC0019040 and No.~DE-SC0024220), NSF QLCI (award No.~OMA-2120757),  NSF STAQ program, AFOSR, AFOSR MURI, and DARPA SAVaNT ADVENT. Support is also acknowledged from the U.S.~Department of Energy, Office of Science, National Quantum Information Science Research Centers, Quantum Systems Accelerator.
L.~T.~B. acknowledges support from the DARPA Quantum Benchmarking program under IAA 8839, Annex 130.
Y.K.L. acknowledges support from the National Institute of Standards and Technology.

%\clearpage
%\bibliographystyle{vancouver}
\bibliography{references}

%\clearpage
%\newpage
\onecolumngrid
%\section*{Supplemental Material}
%\setcounter{secnumdepth}{1}
%\renewcommand{\thesection}{S\arabic{section}}
%\setcounter{equation}{0}
%\renewcommand{\theequation}{S\arabic{equation}}

\section*{APPENDIX}
\setcounter{secnumdepth}{1}

\renewcommand{\thesection}{A\arabic{section}}

\setcounter{section}{0}
\setcounter{equation}{0}
\renewcommand{\theequation}{A\arabic{equation}}

\noindent Appendix~\ref{app-main} --- Derivation of an algorithm's fragility $\robust_Q$ against perturbative noise. We prove Eq.~\eqref{eq:ResultRobustness}.
\vspace{7pt}

\noindent Appendix~\ref{app-incoherent} --- Description of incoherent noise in terms of perturbative coherent noise. We prove Eq.~\eqref{eq:incoherentoverlap}.
\vspace{7pt}

\noindent Appendix~\ref{app-biased} --- We derive the noise model used in Fig.~\ref{fig:fig2}, which allows interpolating between phase and bit-flip noises. 
\vspace{7pt}

\noindent Appendix~\ref{app-analog} --- Derivation of an analog algorithm's fragility against noise. We prove Eq.~\eqref{eq:meanResultRobustnessCont}.
\vspace{7pt}

\noindent Appendix~\ref{app-simpleexample} --- Simple example of a state-preparation protocol and the noise-resilience of different compilations.
\vspace{7pt}

\noindent  Appendix~\ref{app:tradeoff} --- We prove that there exists a tradeoff relation between an algorithm's resilience and its number of gates or runtime. We derive Eq.~\eqref{eq:tradeoffState}.
\vspace{7pt}

\noindent Appendix~\ref{app:p-spin} --- Resilience and the path length of an optimized compilation for the $p$-spin model.
\vspace{7pt}

\noindent Appendix~\ref{app:frag-expectation} --- Focuses on the fragility of the expectation value of cost functions.

\section{Proof --- State-resilience of a quantum algorithm}
\label{app-main}

In this section, we derive an algorithm's fragility $\robust_Q$ against perturbative coherent noise. We prove Eq.~\eqref{eq:ResultRobustness} in the main text.

To leading order in the perturbation parameters $\theta_l^q$, the squared Bures distance between $\ket{\Psi_D}$ and $\ket{\delta \Psi_D}$ is~\cite{liu2019quantum}
\begin{align}
\label{eq-app:DistanceMetric}
  \robust_Q &\coloneqq \nobreak  2 \big(1\nobreak -\nobreak \big|\langle \psi_D\! \ket{\delta\psi_D}\!\big| \big) = \dist^2\left(\ket{\psi_D}\!, \ket{\delta\psi_D}\right) \approx \frac{1}{4}\sum_{j,k = 1}^D \sum_{q = 1}^{\mathcal{N}_j}\sum_{r = 1}^{\mathcal{N}_k} \mathcal I_F\left(\delta \theta_j^q,\delta \theta_k^r\right) \, \delta\theta_j^q  \delta\theta_k^r
\end{align}
Here, $\mathcal I_F\left(\delta \theta_j^q,\delta \theta_k^r\right)$ is the quantum Fisher information matrix, given by
\begin{align}
    \mathcal{I}_F\left(\delta \theta_j^q,\delta \theta_k^r\right) = 4 \textnormal{Re} \Big( \langle \partial_{\delta \theta_j^q}  \delta\psi_D \ket{\partial_{\delta \theta_k^r}  \delta\psi_D} - \langle \partial_{\delta \theta_j^q}  \delta\psi_D \ket{\delta\psi_D} \langle   \delta\psi_D \ket{\partial_{\delta \theta_k^r}  \delta\psi_D} \Big) \bigg\vert_{\vect{\delta \theta = 0}} \,,
\end{align}
for a pure state $\ket{\delta\psi_D}$ parametrized by $\delta\theta_l^q$~\cite{liu2019quantum}. The Fisher information matrix is evaluated at the unperturbed state, with $\vect{\delta \theta} = \vect{0} $. We denote vectors by bold symbols, $\vect{\delta \theta} \coloneqq \left( \delta \theta_1^1,\delta \theta_1^2,\dots, \delta \theta_l^q,\dots,\delta \theta_D^{\mathcal{N}_D} \right)$, and derivatives of a state by $\ket{\partial_{\delta \theta_j^q} \delta\psi_D} \coloneqq \frac{\partial}{\partial \delta\theta_j^q  } \ket{\delta\psi_D }$.

The state that results from a noisy computation of depth $D$ is
\begin{align}
\label{eq-app:perturbedstate}
    \ket{\delta\psi_D} = \prod_{l=1}^D \bigotimes_{q=1}^{\mathcal{N}_l} e^{-i \delta \theta_l^q Q_l^q} M_l^q V_l^q \ket{\psi_0}.
\end{align}

We denote the gate that evolves the state in the unperturbed algorithm between steps $n$ and $m$ by 
$U_{n ; m} \coloneqq \prod_{l=n}^m \bigotimes_{q=1}^{\mathcal{N}_l}  M_l^q V_l^q$. 
Differentiating Eq.~\eqref{eq-app:perturbedstate} then gives
\begin{align}
    \label{eq-app:Statederiv} \ket{\partial_{\delta\theta_j^q}  \delta\psi_D } \at &=  \prod_{m=j+1}^D \bigotimes_{q=1}^{\mathcal{N}_m} e^{-i \delta \theta_m^q Q_m^q} M_m^q V_m^q \big( -i Q_j^q \big) \prod_{l=1}^j \bigotimes_{r=1}^{\mathcal{N}_l} e^{-i \delta\theta_l^r Q_l^r} M_l^r V_l^r \ket{\psi_0} \at \nonumber \\
    &= -i U_{j+1 ; L} Q_j^q U_{1;j} \ket{\psi_0} = -i U_{j+1;L}  U_{1;j} U_{1;j}^\dag Q_j^q U_{1;j} \ket{\psi_0} \nonumber \\
    &= -i U_{1;L} Q_j^q(t_j)  \ket{\psi_0},
\end{align}
where 
\begin{align}
    Q_j^q(t_j) \coloneqq U_{1;j}^\dag Q_j^q U_{1;j}
\end{align}
is evolved in the Heisenberg picture under the unperturbed algorithm. 
Therefore 
\begin{align}
    \langle \partial_{\delta\theta_j^q}  \delta\psi_D \ket{\partial_{\delta\theta_k^r}  \delta\psi_D} \at &=  \bra{\psi_0} Q_j^q(t_j) Q_k^r(t_k)\ket{\psi_0}  \qquad\qquad\qquad\qquad \textnormal{and}   \\
    \langle   \delta\psi_D \ket{\partial_{\delta\theta_k^r}  \delta\psi_D} \at & =  -i \bra{\psi_0} U_{1;L}^\dag   U_{1;L} Q_k^r(t_k)  \ket{\psi_0}  = -i \bra{\psi_0}  Q_k^r(t_k) \ket{\psi_0},
\end{align}
which leads to
\begin{align}
\label{eq-app:FisherAux1}
    \mathcal I_F \left( \delta \theta_j^q, \delta \theta_k^r \right) &= 4 \textnormal{Re} \Big( \bra{\psi_0} Q_j^q(t_j) Q_k^r(t_k)\ket{\psi_0} -  \bra{\psi_0}  Q^q_j(t_j) \ket{\psi_0} \bra{\psi_0}  Q_k^r(t_k) \ket{\psi_0} \Big) \nonumber \\
    &= 2 \Big( \bra{\psi_0} Q_j^q(t_j) Q_k^r(t_k)\ket{\psi_0} + \bra{\psi_0} Q_k^r(t_k) Q_j^q(t_j)\ket{\psi_0} - 2\bra{\psi_0}  Q_j^q(t_j) \ket{\psi_0} \bra{\psi_0}  Q_k^r(t_k) \ket{\psi_0} \Big)  \nonumber \\
    &= 4 \, \cov_{\ket{\psi_0}} \Big( Q_j^q(t_j),Q_k^r(t_k) \Big).
\end{align}
Inserting~\eqref{eq-app:FisherAux1} into~\eqref{eq-app:DistanceMetric} yields
\begin{align}
\label{eq-app:ResultRobustness}
\robust_Q &\approx \sum_{j,k = 1}^D \sum_{q = 1}^{\mathcal{N}_j}\sum_{r = 1}^{\mathcal{N}_k}  \cov_{\ket{\psi_0}} \Big( Q_j^q(t_j), Q_k^r(t_k) \Big) \, \delta\theta_j^q \delta\theta_k^r,
\end{align}
which completes the proof of Eq.~\eqref{eq:ResultRobustness} in the main text.

The approximation error in Eq.~\eqref{eq-app:ResultRobustness} can be bounded by the remainder in a third-order multivariate Taylor expansion of $|\!\bra{\psi_D} \delta \psi_D \rangle|$. Following Ref.~\cite{folland2002advanced}, the difference $R$ between the right hand side and left hand side of Eq.~\eqref{eq-app:ResultRobustness} 
satisfies 
\begin{align}
    |R| &\leq \frac{M}{6} \left( \sum_{l=1}^{D} \sum_{q=1}^{\mathcal{N}_l}|\delta \theta_l^q|  \right)^3
\end{align}
where $M \coloneqq \max \left| \frac{\partial^{3}}{\partial \vect{\delta \theta}} \big( |\!\bra{\psi_D} \delta \psi_D \rangle| \big) \right|$ and $\frac{\partial^{3}}{\partial \vect{\delta \theta}}$ denotes any third-order partial derivative with respect to the $\delta \theta$s.

\section{Proof --- resilience against decoherent noises}
\label{app-incoherent}

In this section, we show how to model small amounts of decoherence, including dephasing and depolarizing noise, by averaging over perturbative stochastic coherent errors. We also prove Eq.~\eqref{eq:incoherentoverlap} in the main text, which holds for general uncorrelated perturbative noise processes.

We begin by considering dephasing noise. The density matrix $\rho$ of a qubit that undergoes a dephasing error with probability $p$ along the $z$ direction undergoes an evolution~\cite{preskill2015lecture}
\begin{align}
    \label{eq-app:dephasing}
    \rho \longrightarrow \rho' = \left( 1-\frac{p}{2} \right)\rho + \frac{p}{2} \left( Z \rho Z \right).
\end{align}
$X$, $Y$, and $Z$ denote the Pauli matrices.
Compare this to the evolution under the noisy coherent channel $e^{-i \delta \theta Z}$:
\begin{align}
    e^{-i \delta \theta Z} \rho e^{i \delta \theta Z}  &=  \cos^2(\delta \theta)  \rho - i \,  \sin(\delta \theta) \cos( \delta \theta ) [Z,\rho] +  \sin^2(\delta \theta)  Z \rho Z.  
\end{align}
Averaging over the noise, we get that
\begin{align}
\label{eq-app:dephasing2}
    \overline{ e^{-i \delta \theta Z} \rho e^{i \delta \theta Z} } = \overline{ \cos^2(\delta \theta) } \rho  + \overline{ \sin^2(\delta \theta) } Z \rho Z   &= \left( \overline{ 1- \delta \theta^2} \right) \rho + \overline{\delta \theta^2} Z \rho Z  \nonumber \\
    &= \left( 1-\sigma^2 \right) \rho + \sigma^2 Z \rho Z, 
\end{align}
to leading order in $\delta \theta$. In the first line, we assumed that the probability of $\delta \theta$ is equal to that of $-\delta \theta$, which implies $\overline{\sin(\delta \theta) \cos(\delta \theta)} = 0$. In the second line, we assumed a perturbative regime, $\delta \theta \ll 1$, with $\overline{\delta \theta^2} = \sigma^2$.

By comparing Eqs.~\eqref{eq-app:dephasing} and~\eqref{eq-app:dephasing2}, we confirm that the framework introduced in the main text can describe dephasing errors that occur with a small probability $p = 2\sigma^2$, where $\overline{ \delta \theta^2 } = \sigma^2$ and $Q = Z$.

Similar calculations show that averaging over coherent noises can model depolarizing noise~\cite{preskill2015lecture},
\begin{align}
    \label{eq-app:depolarizing}
    \rho \longrightarrow \rho' = (1-p)\rho + \frac{p}{3} \left( X \rho X + Y \rho Y +  Z \rho Z \right),
\end{align}
by assuming that simultaneous (uncorrelated) error channels $Q_{x} = X$, $Q_{y} = Y$, and $Q_{z} = Z$ affect a qubit. That is, we consider a noisy coherent channel:
\begin{equation}
 \rho'=e^{-i(\delta\theta_x X+\delta\theta_yY+\delta\theta_z Z)}\rho e^{i(\delta\theta_x X+\delta\theta_yY+\delta\theta_z Z)},
\end{equation}
with $\overline{\delta\theta_i}=0$, $\overline{\delta\theta_i\delta\theta_j}=0$ and $\overline{\delta\theta_i^2}=\sigma^2\ll 1$ for all $i,j\in\{x,y,z\}$.

For generic coherent noise processes, after averaging over noise that acts on a pure state $\rho:=\ket{\psi}\!\bra{\psi}$, one obtains a mixed state:
\begin{align}
    \rho' \equiv \overline{  e^{-i \delta \theta Q} \ket{\psi}\!\bra{\psi} e^{i \delta \theta Q}  }.
\end{align}
If $Q$ is a Pauli string satisfying $Q^2 = \id$, and if $\overline{ \sin (\delta \theta)} = 0$ and $\overline{\delta\theta^2}=\sigma^2$, 
\begin{align}
    \rho' \equiv \overline{  e^{-i \delta \theta Q} \ket{\psi}\!\bra{\psi} e^{i \delta \theta Q}  } = (1-\sigma^2) \rho + \sigma^2 Q \rho Q,
\end{align}
which describes a decoherent process along the eigenbasis of $Q$.

Next, we relate an algorithm's average fragility to the overlap between the resulting density matrix $\rho$ from a noisy implementation and the target pure state $\ket{\psi_D}$.  
From the definition of the fragility, $1 - \tfrac{1}{2}\robust_Q \nobreak = \nobreak \left|\langle \psi_D\! \ket{\delta \psi_D}\right|$.
Then, Eq.~\eqref{eq-app:ResultRobustness} implies
\begin{align}
\label{eq-app:openaux}
    \left|\langle \psi_D\! \ket{\delta \psi_D}\right|^2 &=  \left( 1 - \tfrac{1}{2}\robust_Q \right)^2 = 1 - \robust_Q + \frac{1}{4} \robust_Q^2 \nonumber \\
    &\approx 1 -\sum_{j,k = 1}^D \sum_{q = 1}^{\mathcal{N}_j}\sum_{r = 1}^{\mathcal{N}_k}  \cov_{\ket{\psi_0}} \Big( Q_j^q(t_j), Q_k^r(t_k) \Big) \, \delta\theta_j^q \delta\theta_k^r,
\end{align}
to leading order in $\delta \theta$. 
 
Let $\rho_D \coloneqq \overline{\ket{\delta \psi_D}\!\bra{\delta \psi_D}}$ be the noise-averaged final state of the system for uncorrelated noises. Eq.~\eqref{eq-app:openaux} becomes
\begin{align}
    \tr{\rho_D \ket{\psi_D}\!\bra{\psi_D}} \approx 1 - \sum_{l= 1}^D \sum_{q = 1}^{\mathcal{N}_l}  \sigma_{lq}^2 \,\var_{\ket{\psi_{l}}} \big(Q_l^q\big).
\end{align}
This expression quantifies the effect of an incoherent perturbative noise process on the system's state.
 This proves Eq.~\eqref{eq:incoherentoverlap} in the main text.

\section{A model for biased noise}
\label{app-biased}
In this section, we derive a noise model that allows interpolating between dephasing and bitflip channels. We use this model to simulate the noise resilience of error-detecting codes in Fig.~\ref{fig:fig2}.

We construct a noise model that i) allows for interpolation between the relative strength of noise in each 
Pauli basis, and ii) can be mapped to a coherent error, as detailed in Appendix~\ref{app-incoherent}. 
The latter condition restricts us to considering channels of the form
\begin{align}
        \rho \longrightarrow \rho' = \left((1-p_x) \left(\cdot\right) + p_x  X \left(\cdot\right) X \right) \circ \left((1-p_y) \left(\cdot\right) + p_y Y \left(\cdot\right) Y \right) \circ \left((1-p_z) \left(\cdot\right) + p_z  Z \left(\cdot\right) Z \right)\left(\rho\right),
\end{align}
where $\circ$ denotes channel composition. 
The former condition can then be met by the channel
\begin{align} \label{eq-app:bias}
        \rho \longrightarrow \rho' = \left((1-p_x) \left(\cdot\right) + p_x  X \left(\cdot\right) X \right) \circ \left((1-p_z) \left(\cdot\right) + p_z  Z \left(\cdot\right)  Z \right)\left(\rho\right),
\end{align}
with total error rate $p = p_x + p_z$ and a bias toward Pauli X errors $\eta_X$. 
We can then write $p_x = \eta_x p$ and $p_z = (1 - \eta_x) p$.
At $\eta_x = 0$ we have $p_x = 0$, $p_z = p$ and Eq.~(\ref{eq-app:bias}) reduces to a dephasing channel
\begin{align} 
        \rho \longrightarrow \rho' = (1-p)\rho + p Z \rho Z.
\end{align}
At $\eta_x = 0.5$ we have $p_x = p / 2$, $p_z = p / 2$ and Eq.~(\ref{eq-app:bias}) describes a channel with an equal probability of bit and phase flip errors 
\begin{align}
        \rho \longrightarrow \rho' = \left(1-\frac{p}{2}\right)^2 \rho + \left(1 - \frac{p}{2}\right) \frac{p}{2} \left( X \rho X + Z \rho Z \right) + \frac{p^2}{4} Y \rho Y.
\end{align}
At $\eta_x$ = 1, Eq.~(\ref{eq-app:bias}) tends to a bitflip channel
\begin{align} 
        \rho \longrightarrow \rho' = (1-p)\rho + p  X \rho X.
\end{align}

In our numerical simulations we consider coherent noise.
We map the above noise model to coherent noise using the procedure detailed in Section \ref{subsec:incohe}. 
We assume the same noise acts after every gate and at each idling step.
In simulation this assumption amounts to applying one or more coherent errors based on the above error channel to 
every qubit after each layer of parallel operations. 
The rotation angles of these coherent errors are random and independent but drawn from the same distribution, with a width  determined by $p$.

\section{Proof --- Resilience of analog quantum algorithms}
\label{app-analog}

In this section, we study the fragility of analog quantum algorithms. In particular, we prove Eq.~\eqref{eq:meanResultRobustnessCont} in the main text.

In an analog algorithm the sequence of gates $V_l$ is replaced by a time-dependent Hamiltonian $H(t)$. We assume a Hamiltonian that acts across all qubits, so the index $q = \{1,\ldots,\mathcal{N}_l\}$ is no longer necessary. 
We start from Eq.~\eqref{eq-app:ResultRobustness}.
To take a continuum limit, we assume $\delta \theta_l \longrightarrow dW_t$, where $dW_t$ is the noise affecting the system over an infinitesimal duration $dt$. That is, instead of the stochastic phases $\delta \theta_l$ we have stochastic functions $dW_t$. In a time-continuous Hamiltonian description, the ideal dynamics $e^{-i H_t dt}$ is perturbed to $e^{ -i H_t dt-i Q dW_t } = e^{-i (H_t + \xi_t Q) dt}$, where we have parameterized $dW_t=\xi_t dt$, using a continuous noise function $\xi_t$.
The time-continuous version of Eq.~\eqref{eq-app:ResultRobustness} then is given by the It\^{o} stochastic integral~\cite{jacobs2010stochastic},
\begin{align}
\label{eq-app:ResultRobustnessACont}
\robust_Q &\approx \int_0^T  \cov_{\ket{\psi_0}} \Big( Q(t), Q(t') \Big) \, dW_t  \, dW_{t'} \nonumber \\
&= \int_0^T  \cov_{\ket{\psi_0}} \Big( Q(t), Q(t') \Big) \, \xi_t  \, \xi_{t'} \, dt dt'.
\end{align}
$Q(t)$ denote observables in the Heisenberg picture.

If $dW_t$ represents zero-mean white noise, satisfying $\overline{dW_t dW_{t'}} = \delta(t-t') \gamma_t dt$ (i.e. $W_t$ is a Wiener process), the noise-averaged fragility is
\begin{align}
\label{eq-app:meanResultRobustnessACont}
\overline{\robust_Q} &\approx \int_0^T \gamma_t \,  \var_{\ket{\psi_t}} \big( Q \big) \,  dt,  
\end{align}
The function $\gamma_t$ dictates the magnitude of the noises at time $t$.

\section{Example --- Different compilations of a state transformation and their resilience}
\label{app-simpleexample}

In this section, we consider a simple example of a state-preparation protocol. We find compilations with different noise resilience.

Consider a two-qubit unitary that transforms $\ket{00}$ to $\ket{11}$.  We consider two implementations of such transformation and show that one of the implementations can be more resilient against noise while being sensitive to another noise. % \cite{GyhmRosa}.

Consider $V_a(t) := e^{-iH_a t}$ generated by the Hamiltonian
\begin{equation}
    H_a \coloneqq -i\Big(\ketbra{00}{11}-\ketbra{11}{00}\Big).
\end{equation}
At time $t=\frac{\pi}{2}$, the state reaches the target $\ket{11}$ i.e.,
\begin{equation}
    V_a\left(\frac{\pi}{2}\right) \ket{00} = e^{-iH_a\frac{\pi}{2}}\ket{00} = \ket{11}.
\end{equation}

Alternatively, the Hamiltonian
\begin{equation}
    H_b \coloneqq -\frac{i}{2}\Big(\big(\ketbra{0}{1}-\ketbra{1}{0}\big)\otimes\mathbb{I}+\mathbb{I}\otimes\big(\ketbra{0}{1}-\ketbra{1}{0}\big)\Big) = -\frac{1}{2} \left( Y \otimes \id + \id \otimes Y \right)
\end{equation}
can perform the same state transfer. We denote 
$V_b(t):=e^{-iH_bt}$, which satisfies
\begin{equation}
    V_b(\pi)\ket{00} = \ket{11}.
\end{equation}

Under each compilation, the evolved state is
\begin{equation}
    \ket{\psi_{a,t}} := e^{-iH_at}\ket{00} = \cos(t)\ket{00}+\sin(t)\ket{11},
\end{equation}
and
\begin{equation}
    \ket{\psi_{b,t}} := e^{-iH_bt}\ket{00} = \cos^2\left(\frac{t}{2}\right)\ket{00}+\frac{1}{2}\sin(t)\ket{01}+\frac{1}{2}\sin(t)\ket{10}+\sin^2\left(\frac{t}{2}\right)\ket{11}.
\end{equation}

Let us compare the resilience of both implementations to different noises. 
We consider two noise operators
\begin{align}
    Q_{i} &\coloneqq X \otimes X, \qquad\qquad \textnormal{and} \\
    Q_{ii} &\coloneqq \frac{1}{4}\left(\mathbb{I}-Z \right)\otimes\left(\mathbb{I}-X\right).
\end{align}
The fragility against noise is given by Eq.~\eqref{eq:meanResultRobustnessCont} in the main text: 
\begin{equation}
\label{eq:App_1Fragility_1}
    \overline{\robust_Q} \approx \gamma \int_0^T \var_{\ket{\psi_t}} \big( Q \big) \,  dt,
\end{equation}
where we assumed that $\gamma_t \equiv \gamma$ is constant for simplicity.

Let us calculate the fragility when the dynamics are governed by the Hamiltonian $H_a$. 
%To evaluate fragility given in Eq. \eqref{eq:App_1Fragility_1} for noise observable $Q_i$ and $Q_{ii}$, we proceed by calculating 
%\begin{equation}
    %\ket{\psi_{a,t}} := e^{-iH_at}\ket{00} = \cos(t)\ket{00}+\sin(t)\ket{11}.
%\end{equation}
The variance of the error operator $Q_{i}$ and $Q_{ii}$ with respect to $\ket{\psi_{a,t}}$ is 
\begin{eqnarray}\label{App_var_11_Eq}
    \var_{\ket{\psi_{a,t}}} \big( Q_i \big) &=& \bra{\psi_{a,t}}Q_i^2\ket{\psi_{a,t}}-\left(\bra{\psi_{a,t}}Q_i\ket{\psi_{a,t}}\right)^2 = \cos^2(2t),\\
    \var_{\ket{\psi_{a,t}}} \big( Q_{ii} \big) &=& \bra{\psi_{a,t}}Q_{ii}^2\ket{\psi_{a,t}}-\left(\bra{\psi_{a,t}}Q_{ii}\ket{\psi_{a,t}}\right)^2= \frac{1}{8}\sin ^2(t) (\cos (2 t)+3).
\end{eqnarray}
Employing the expression of $\var_{\ket{\psi_{a,t}}}\big( Q_i \big)$ and $\var_{\ket{\psi_{a,t}}} \big( Q_{ii}\big)$ we can calculate fragility given in Eq. \eqref{eq:App_1Fragility_1} as 

\begin{eqnarray}
    \overline{\robust_{\ket{\psi_{a,t}}}^{Q_i}} &\approx& \int_0^{T=\frac{\pi}{2}}  \, \cos^2(2t)  \,  dt = \frac{\pi}{4},\\
    \overline{\robust_{\ket{\psi_{a,t}}}^{Q_{ii}}} &\approx& \int_0^{T=\frac{\pi}{2}}  \, \frac{1}{8} \sin ^2(t) (\cos (2 t)+3)  \,  dt = \frac{5\pi}{64}.
\end{eqnarray}

Next, we evaluate the fragility when the dynamics are governed by the Hamiltonian $H_b$. %Similarly as before, we can calculate 
%\begin{equation}
%    \ket{\psi_{b,t}} := e^{-iH_bt}\ket{00} = \cos^2\left(\frac{t}{2}\right)\ket{00}+\frac{1}{2}\sin(t)\ket{01}+\frac{1}{2}\sin(t)\ket{10}+\sin^2\left(\frac{t}{2}\right)\ket{11}.
%\end{equation}
The variances of the error operators $Q_i$ and $Q_{ii}$ with respect to  $\ket{\psi_{b,t}}$ are
\begin{eqnarray}\label{App_var_22_Eq}
    \var_{\ket{\psi_{b,t}}} \big( Q_{i} \big) &=& \bra{\psi_{b,t}}Q_{i}^2\ket{\psi_{b,t}}-\left(\bra{\psi_{b,t}}Q_{i}\ket{\psi_{b,t}}\right)^2 = \frac{1}{2}\cos ^2(t)\big(3-\cos (2 t)\big),\\
    \var_{\ket{\psi_{b,t}}} \big( Q_{ii} \big) &=& \bra{\psi_{b,t}}Q_{ii}^2\ket{\psi_{b,t}}-\left(\bra{\psi_{b,t}}Q_{ii}\ket{\psi_{b,t}}\right)^2=\frac{1}{64} \Big(\sin (t)+\cos (t)-1\Big)^2 \Big(6+2\cos(t)+2\sin(t)-\sin(2t)\Big)\nonumber\\.
\end{eqnarray}
Employing the expression of $\var_{\ket{\psi_{b,t}}}\big( Q_{i} \big)$ and $\var_{\ket{\psi_{b,t}}} \big( Q_{ii}\big)$ we can calculate fragility given in Eq. \eqref{eq:App_1Fragility_1} as 
\begin{eqnarray}
    \overline{\robust_{\ket{\psi_{b,t}}}^{Q_i}} &\approx& \int_0^{T=\pi}  \, \frac{1}{2}\cos ^2(t)\big(3-\cos (2 t)\big)  \,  dt = \frac{5 \pi }{8},\\
    \overline{\robust_{\ket{\psi_{b,t}}}^{Q_{ii}}} &\approx& \int_0^{T=\pi}  \, \frac{1}{64} \Big(\sin (t)+\cos (t)-1\Big)^2 \Big(6+2\cos(t)+2\sin(t)-\sin(2t)\Big)  \,  dt = \frac{15 \pi }{128}-\frac{1}{6}.
\end{eqnarray}
Now, note that 
\begin{eqnarray}
    \frac{5\pi}{8}\approx\overline{\robust_{\ket{\psi_{b,t}}}^{Q_{i}}} &>&\overline{\robust_{\ket{\psi_{a,t}}}^{Q_{i}}}\approx \frac{2\pi}{8}\\
    0.201 \approx \frac{15 \pi }{128}-\frac{1}{6}\approx\overline{\robust_{\ket{\psi_{b,t}}}^{Q_{ii}}} &<&\overline{\robust_{\ket{\psi_{a,t}}}^{Q_{ii}}}\approx\frac{5\pi}{64}\approx 0.245.
\end{eqnarray}
Thus, we can say the evolution from $\ket{00}$ to $\ket{11}$ generated by the Hamiltonian $H_b$ is more fragile than the evolution from $\ket{00}$ to $\ket{11}$ generated by the Hamiltonian $H_a$ when it is subjected to noise $Q_{i}$. On the other hand, the evolution from $\ket{00}$ to $\ket{11}$ generated by the Hamiltonian $H_a$ is more fragile than the evolution from $\ket{00}$ to $\ket{11}$ generated by the Hamiltonian $H_b$ when it is subjected to noise $Q_{ii}$.

\section{Proof --- resilience-runtime tradeoff relations}
\label{app:tradeoff}

In this section we prove Eq.~\eqref{eq:tradeoffState} in the main text.

In the main text, using \cref{eq:ResultRobustness} (proved in \cref{app-main}), we showed that the average fragility of an algorithm subject to uncorrelated noise is
\begin{align}
\overline{\robust_Q} &\approx \sum_{l= 1}^D \sum_{q = 1}^{\mathcal{N}_l}  \var_{\ket{\psi_{l}}} \big(Q_l^q\big) \sigma_{lq}^2. 
\end{align}
We also defined, in \cref{eq:distance}, the geometric quantity 
\begin{align}
   \length_Q \coloneqq \sum_{l q} |\theta_l^q| \sqrt{\var_{\ket{\psi_l}} (Q_l^q)}.
\end{align}
The two can be related by using the Cauchy-Schwarz inequality:
\begin{align}
    \min_{l',q'} \left( \frac{\sigma_{l' q'}}{\theta_{l'}^{q'}} \right)^2 \length_Q^2 &= \min_{l',q'} \left( \frac{\sigma_{l' q'}}{\theta_{l'}^{q'}} \right)^2 \left( \sum_{l q} |\theta_{l}^{q}| \sqrt{\var_{\ket{\psi_l}} (Q_l^q)} \right)^2 \leq \min_{l',q'} \left( \frac{\sigma_{l' q'}}{\theta_{l'}^{q'}} \right)^2 \sum_{l q} (\theta_l^q)^2 \, \var_{\ket{\psi_l}} (Q_l^q) \sum_{l q} 1 \nonumber \\
    &\leq  N_G  \sum_{l q} \left( \frac{\sigma_{l q}}{\theta_l^q} \right)^2 (\theta_l^q)^2 \, \var_{\ket{\psi_l}} (Q_l^q) = N_G  \sum_{l q}   \sigma_{l q}^2 \, \var_{\ket{\psi_l}} (Q_l^q) \nonumber \\
    &\approx N_G \, \overline{\robust_Q},
\end{align}
where we use that the total number of gates in the circuit is $N_G \coloneqq \sum_{l=1}^D \sum_{q = 1}^{\mathcal{N}_l} 1 = \sum_{l=1}^D\mathcal{N}_l$. 
This proves Eq.~\eqref{eq:tradeoffStateCircuit} in the main text.

A similar constraint holds for analog algorithms. Via \cref{eq:meanResultRobustnessCont} the average fragility of an analog algorithm is
\begin{align}
    \overline{\robust_Q} &\approx \int_0^T \gamma_t \,  \var_{\ket{\psi_t}} \big( Q \big) \,  dt,
\end{align} 
and $\mathcal{L}_Q \equiv \int_0^T \var_{\ket{\psi_t}} (Q) dt$. 
The Cauchy-Schwarz inequality for functions implies the time-continuous version of Eq.~\eqref{eq:tradeoffStateCircuit} given in \cref{eq:tradeoffStateCont}:
\begin{align}
    T \overline{\robust_Q} &\geq \min_t \gamma_t \, \mathcal{L}_Q^2.
\end{align}

\section{Resilience of a time-optimized compilation for the $p$-spin model}
\label{app:p-spin}
We study the resilience and the path length of an optimized compilation for the $p$-spin model.

The ferromagnetic $p$-spin model is described by the time-dependent Hamiltonian $H_t \nobreak=\nobreak (1 \nobreak-\nobreak g_t) H_0 \nobreak+\nobreak g_t H_1$, where 
%\begin{equation}
%    H_0 = \frac{n}{2}\left(\mathbb{I}-\frac{M_x}{n}\right), \quad H_1 = \frac{n}{2}\left(\mathbb{I}-\frac{M_z^p}{n^p}\right),
%\end{equation}
\begin{equation}
    H_0 = - \frac{1}{2} M_x, \qquad H_1 =  -\frac{n}{2} \frac{M_z^p}{n^p}.
\end{equation}
%\begin{align}
%H_0 = -  \sum_{j=1}^n X_j/2 \qquad \textnormal{and} \quad H_1 = - \big( \sum_{j=1}^n Z_j \big)^p/(2 n^{p-1})
%\end{align}
$M_x = \sum_{q=1}^n X_q$ and $M_{z} = \sum_{q=1}^n Z_q$ are the total magnetization of the $n$ qubits 
%where $X$ and $Z$ are the Pauli matrices 
along the $x$ and $z$ directions, respectively. 
For odd $p$ and $n$, a time-optimized protocol reaches the ground state of $H_1$ in polynomial time with unit fidelity. More precisely, we drive the system from the ground state of $H_0$ using first the Hamiltonian $H_1$ for time $t_1$ then the Hamiltonian $H_0$ for time $t_2$ using time evolution unitaries defined by
\begin{equation}
\label{QAOA_t1_t2}
    e^{-it_1H_1} = e^{i\frac{\pi}{4} M_z^p} \quad e^{it_2H_0} = e^{-i\frac{\pi}{4} M_x} \Rightarrow\;\; t_1 = \frac{\pi}{8}n^{p-1}\quad\text{and}\quad t_2 = \frac{\pi}{2}.
\end{equation}
This procedure perfectly transfers the system from the ground state of $H_0$ to the ground state of $H_1$~\cite{pspinQAOA}.
The system's initial, intermediate, and final states are
\begin{equation}
    \ket{\psi_0} = \ket{+}^{\otimes n} = \frac{1}{2^{n/2}} \left(\ket{0}+\ket{1}\right)^{\otimes n},  \quad\qquad \ket{\psi_1} = e^{-iH_1t_1}\ket{\psi_0}, 
    \quad\qquad\text{and}\quad\qquad \ket{\psi_D} = e^{-iH_0t_2} e^{-iH_1t_1}\ket{\psi_0}.
\end{equation}

The length $\mathcal{L} = \int_0^T \!\sqrt{ \var_{\psi_t} (H_t) } dt$ of the path traveled by the state is  
\begin{equation}\label{QAOA_path_length}
    \mathcal{L}  = t_1 \sqrt{\var_{\ket{\psi_1}}(H_1)}+t_2 \sqrt{\var_{\ket{\psi_D}}(H_0)} = t_1 \sqrt{\var_{\ket{\psi_0}}(H_1)}+t_2 \sqrt{\var_{\ket{\psi_1}}(H_0)},
\end{equation}
where we used that unitary dynamics with a Hamiltonian $H$ leaves the Hamiltonian's variance unchanged.
In order to calculate the path length in Eq.~\eqref{QAOA_path_length}, we proceed by calculating $\var_{\ket{\psi_0}}(H_1)$ and $\var_{\ket{\psi_1}}(H_0)$. 

Let us first calculate $\var_{\ket{\psi_0}}(H_1)$:
\begin{eqnarray}\label{QAOA_variance_first_term}
    \var_{\ket{\psi_0}}(H_1) = \bra{\psi_0}H_1^2\ket{\psi_0}-(\bra{\psi_0}H_1\ket{\psi_0})^2 = \frac{n^2}{4}\left(\bra{\psi_0} \frac{M_z^{2p}}{n^{2p}}\ket{\psi_0}-\left(\bra{\psi_0} \frac{M_z^p}{n^p} \ket{\psi_0}\right)^2\right).\nonumber\\
\end{eqnarray}
Thus, to evaluate the above expression we need $\bra{\psi_0}M_z^p\ket{\psi_0}$ and $\bra{\psi_0}M_z^{2p}\ket{\psi_0}$. The first one is
\begin{eqnarray}
    \bra{\psi_0}M_z^{p}\ket{\psi_0} &=& \bra{\psi_0}\left(\sum_{q=1}^n Z_{q}\right)^{p}\ket{\psi_0} = \sum_{\textbf{k}:k_1+k_2+\ldots k_n=p}\binom{p}{k_1\;k_2\;\ldots k_n}\bra{\psi_0}\left(Z^{k_1}\otimes Z^{k_2}\ldots \otimes Z^{k_n}\right)\ket{\psi_0}\nonumber\\
    &=& \sum_{\textbf{k}:k_1+k_2+\ldots k_n=p}\binom{p}{k_1\;k_2\;\ldots k_n}\bra{+}^{\otimes n}\left(Z^{k_1}\otimes Z^{k_2}\ldots \otimes Z^{k_n}\right)\ket{+}^{\otimes n}\nonumber\\
    &=& \sum_{\textbf{k}:k_1+k_2+\ldots k_n=p}\binom{p}{k_1\;k_2\;\ldots k_n}\bra{+}Z^{k_1}\ket{+}\bra{+}Z^{k_2}\ket{+}\ldots  \bra{+}Z^{k_n}\ket{+},
\label{QAOA_sum_intermediate}
\end{eqnarray}
where $\textbf{k}$ is a $n$-dimensional vector whose entries are given as $(k_1\; k_2\;\ldots k_n)$. Note that as $p$ is odd, for each $\textbf{k}$ term in the sum~\eqref{QAOA_sum_intermediate}, there has to be at least one $k_i$ for $i\in\{1,\ldots n\}$, which is odd. Otherwise, $k_i$ for $i\in\{1,\ldots n\}$ will not add up to $p$ which is assumed to be odd. Observing that $\bra{+}Z^{k_j}\ket{+} = 0$ when $k_j$ is odd, we conclude
\begin{equation}\label{QAOA_odd_power}
    \bra{\psi_0}M_z^{p}\ket{\psi_0} = 0.
\end{equation}
Moreover, 
\begin{equation}\label{QAOA_sum_intermediate2}
    \bra{\psi_0}M_z^{2p}\ket{\psi_0} = \bra{\psi_0}\left(\sum_{q=1}^n Z_{q}\right)^{2p}\ket{\psi_0} = \sum_{\textbf{k}:k_1+k_2+\ldots k_n=2p}\binom{2p}{k_1\;k_2\;\ldots k_n}\bra{+}Z^{k_1}\ket{+}\bra{+}Z^{k_2}\ket{+}\ldots  \bra{+}Z^{k_n}\ket{+}.\nonumber\\
\end{equation}
Note that only $\textbf{k}$ whose elements are even and add up to $2p$ contribute to Eq.~\eqref{QAOA_sum_intermediate2}. Therefore, we can write
\begin{equation}\label{QAOA_even_power}
\bra{\psi_0}M_z^{2p}\ket{\psi_0} =\sum_{\substack{\textbf{k}:k_1+k_2+\ldots k_n=2p\\ {s.t }\;\forall i\; k_i \;\text{are even}}} \binom{2p}{k_1\;k_2\;\ldots k_n} = \sum_{i=0}^{2p} \binom{2p}{i}(-n)^{2p-i}\sum_{j=0}^{i}\mathcal{S}(i,j)2^{i-j}(n)_j,
\end{equation}
where  $(n)_j = \frac{n!}{(n-j)!}$  and $\mathcal{S}(i,j)$ is known as the Stirling number of the second kind,  defined by
\begin{equation}
    \mathcal{S}(i,j) \coloneqq \frac{1}{j!}\sum_{l=0}^j (-1)^{j-l}\binom{j}{l}l^i = \sum_{l=0}^{j}\frac{(-1)^{j-l}l^i}{(j-l)! \,l!}.
\end{equation}
%and 
%\begin{equation}
%    (n)_j = \frac{n!}{(n-j)!}.
%\end{equation}

Substituting the expression of $\bra{\psi_0}M_z^{2p}\ket{\psi_0}$ from Eq.~\eqref{QAOA_even_power} in Eq.~\eqref{QAOA_variance_first_term}, and using that  $\bra{\psi_0}M_z^{p}\ket{\psi_0} = 0$ as obtained in Eq.~\eqref{QAOA_odd_power}, we get
\begin{equation}\label{QAOA_variance_first_term_final}
    \var_{\ket{\psi_0}}(H_1) = \frac{n^{2-2p}}{4} \bra{\psi_0}M_z^{2p}\ket{\psi_0} = \frac{n^{2-2p}}{4} \sum_{i=0}^{2p} \binom{2p}{i}(-n)^{2p-i}\sum_{j=0}^{i}\mathcal{S}(i,j)2^{i-j}(n)_j.
\end{equation}

Next, we calculate the term $\var_{\ket{\psi_1}}(H_0)$ in Eq.~\eqref{QAOA_path_length}. To calculate it, we first calculate 
\begin{align}
    \ket{\psi_1} &= e^{-it_1H_1}\ket{\psi_0} = e^{i\frac{\pi}{4}M_z^p}\ket{\psi_0} = e^{i\frac{\pi}{4}M_z}\ket{\psi_0} = \left(e^{i\frac{\pi}{4}Z}\ket{+}\right)^{\otimes n} = \left(\cos\left(\frac{\pi}{4}\right)\ket{+}+i\sin\left(\frac{\pi}{4}\right)\ket{-}\right)^{\otimes n} \nonumber \\
    &= \frac{1}{2^{n/2}}\left(\ket{+}+i\ket{-}\right)^{\otimes n}.
\end{align}
where we use $e^{i\frac{\pi}{4}M_z^p} = e^{i\frac{\pi}{4}M_z}$ in the third equality. We need to evaluate 
\begin{equation}\label{QAOA_variance_second_term}
    \var_{\ket{\psi_1}}(H_0)= \bra{\psi_1}H_0^2\ket{\psi_1} - \left(\bra{\psi_1}H_0\ket{\psi_1}\right)^2 = \frac{n^2}{4}\bra{\psi_1} \frac{M_x^2}{n^2} \ket{\psi_1} - \frac{n^2}{4}\left(\bra{\psi_1} \frac{M_x}{n} \ket{\psi_1}\right)^2.
\end{equation}
To do so, note that
\begin{equation}\label{QAOA1}
    \bra{\psi_1}M_x\ket{\psi_1} = \frac{1}{2^n}\Big(\left(\bra{+}-i\bra{-}\right)\sigma_x\left(\ket{+}+i\ket{-}\right)\Big)^{\otimes n} = 0,
\end{equation}
and
\begin{equation}\label{QAOA12}
    \bra{\psi_1}M_x^2\ket{\psi_1} = \frac{1}{2^n}\Big(\left(\bra{+}-i\bra{-}\right)X^2\left(\ket{+}+i\ket{-}\right)\Big)^{\otimes n} = \frac{2^n}{2^n}=1.
\end{equation}
Substituting the values of $\bra{\psi_1}M_x\ket{\psi_1}$ and $\bra{\psi_1}M_x^2\ket{\psi_1}$ from Eqs.~\eqref{QAOA1} and~\eqref{QAOA12} in Eq.~\eqref{QAOA_variance_second_term}, we obtain: 
\begin{equation}\label{QAOA_variance_second_term_final}
    \var_{\ket{\psi_1}}(H_0) = \frac{1}{4}.
\end{equation}

Finally, substituting the expressions for the variances  given in  Eqs.~\eqref{QAOA_variance_first_term_final} and~\eqref{QAOA_variance_second_term_final} and the expressions for $t_1$, $t_2$ given in Eq.~\eqref{QAOA_t1_t2} in Eq.~\eqref{QAOA_path_length},  we find that
\begin{eqnarray}
\label{eq-app:distanceQAOA}
    \mathcal{L} &=& \frac{\pi}{16}\left( \sum_{i=0}^{2p} \binom{2p}{i}(-n)^{2p-i}\sum_{j=0}^{i}\mathcal{S}(i,j)2^{i-j}(n)_j\right)^{\frac{1}{2}}+\frac{\pi}{4}.
\end{eqnarray}
We can simplify re-write $\mathcal{L}$ as follows
\begin{eqnarray}
     \mathcal{L} &=& \frac{\pi}{16}\left( \sum_{i=0}^{2p} \binom{2p}{i}(-n)^{2p-i}\sum_{j=0}^{i}\mathcal{S}(i,j)2^{i-j}(n)_j\right)^{\frac{1}{2}}+\frac{\pi}{4}  = \frac{n^p\pi}{16}\left( \sum_{i=0}^{2p} \binom{2p}{i}2^i(-1)^{i}\sum_{j=0}^{i}\frac{\mathcal{S}(i,j)}{n^i}2^{-j}(n)_j\right)^{\frac{1}{2}}+\frac{\pi}{4} \nonumber \\
     &=& \frac{n^p\pi}{16}\left( \sum_{i=0}^{2p} \binom{2p}{i}(-2)^i\sum_{k=0}^{n}\frac{1}{2^n n^i}\binom{n}{k}k^i\right)^{\frac{1}{2}}+\frac{\pi}{4} \label{Eq:A45}\\&=& \frac{n^p\pi}{16}\left( \frac{1}{2^n}\sum_{k=0}^{n}\binom{n}{k}\sum_{i=0}^{2p} \binom{2p}{i}\left(\frac{-2k}{n}\right)^i\right)^{\frac{1}{2}}+\frac{\pi}{4} = \frac{n^p\pi}{16}\left( \frac{1}{2^n}\sum_{k=0}^{n}\binom{n}{k}\left(1-\frac{2k}{n}\right)^{2p}\right)^{\frac{1}{2}}+\frac{\pi}{4} \nonumber \\
     &=& \frac{n^p\pi}{16}\left( \sum_{k=0}^{n}\frac{\binom{n}{k}}{2^n}\left(1-\frac{2k}{n}\right)^{2p}\right)^{\frac{1}{2}}+\frac{\pi}{4},\label{A49}
\end{eqnarray}
where to write the equality in Eq. \eqref{Eq:A45}, we use 
\begin{equation}
\sum_{j=0}^{i}\mathcal{S}(i,j)2^{-j}(n)_j = \frac{1}{2^n}\sum_{k=0}^n\binom{n}{k}k^i.
\end{equation}

\begin{figure}[t]
    \centering
    \includegraphics[width=18cm]{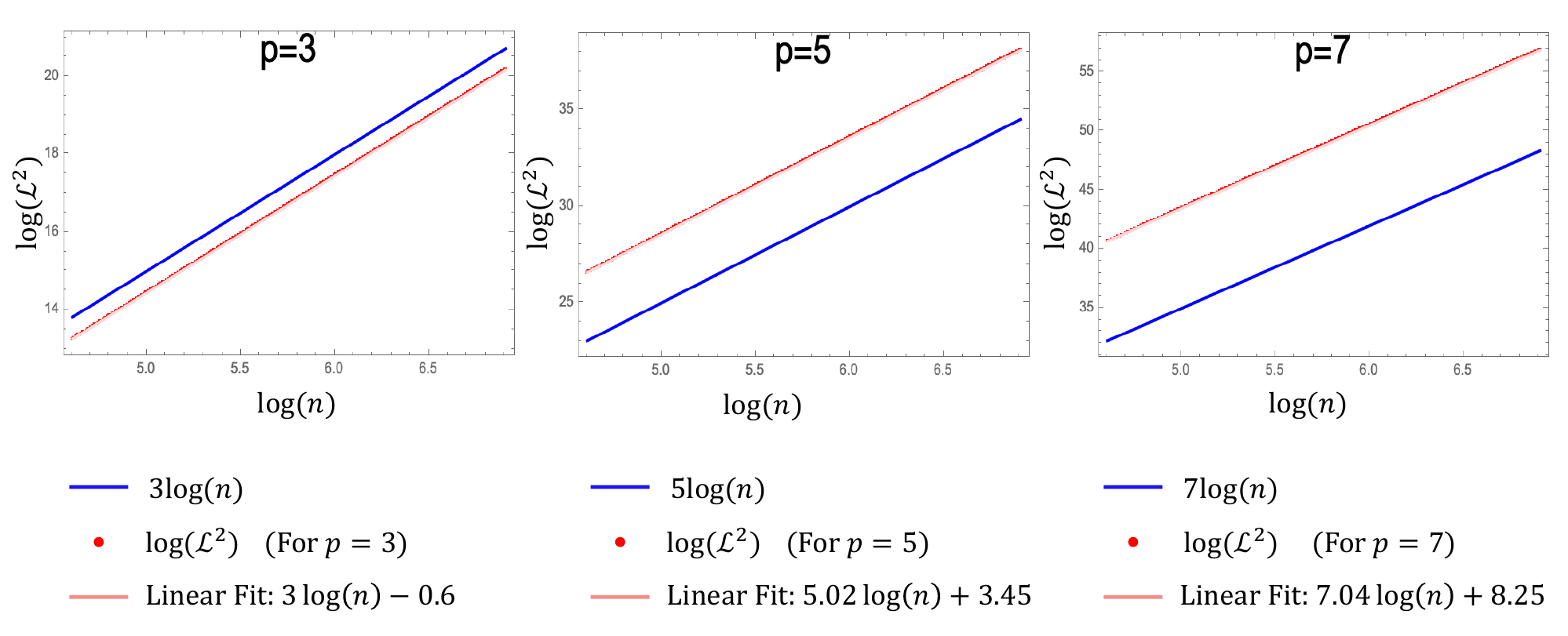}
    \caption{\label{fig5} \textbf{The path length of a time-optimized compilation of the $p$-spin model.}
    Log-log plot of the path length $\mathcal{L}$ versus the number of qubits $n$ for an optimized compilation of the p-spin model. The path length $\mathcal{L}$ is given by Eq.~\eqref{A49}. We plot $\log(\mathcal{L}^2)$ (dotted red) and a linear fit (continuous light red) for $p=\{3,\;5,\;7\}$, confirming a linear scaling with $\log(n)$. To compare, we also plot the function $p\log(n)$ for $p=3,\;5,\;7$ (continuous blue).} 
\end{figure}
In Fig.~\ref{fig5}, we plot and fit the results of Eq.~\eqref{A49} to find that
\begin{equation}
    \log\left(\mathcal{L}^2\right)\sim p\log(n)=\log(n^p)\;\;\Rightarrow\;\; \mathcal{L}^2\sim n^p.
\end{equation}
Using the resilience-runtime tradeoff relation in Eq.~\eqref{eq:tradeoffStateCont}, we obtain
\begin{eqnarray}\label{above234}
    \overline{\robust_Q} &\geq& \min_t \gamma_t \,\, \frac{\mathcal{L}^2}{T} \sim \min_t \gamma_t \,\, \frac{n^p}{T},
\end{eqnarray}
where $T$ can be calculated from Eq. \eqref{QAOA_t1_t2} as
\begin{align}\label{above2341}
    T = t_1 + t_2 = \frac{\pi}{8}n^{p-1} + \frac{\pi}{2}. 
\end{align}
Substituting Eq. \eqref{above2341} into Eq. \eqref{above234}, one can write the following approximate inequality:
\begin{eqnarray}
\label{eq-aux:tradeoffpspin}
    \overline{\robust_Q} &\gtrsim&  \min_t \gamma_t \,\, 
    \frac{n^p}{n^{p-1}}=n\min_t \gamma_t.
\end{eqnarray}
The inequality is almost saturated in this example, since $t_1 \gg t_2$ and $\var_{\ket{\psi_0}}(H_1) \gg \var_{\ket{\psi_D}}(H_0)$. Then, Eq.~\eqref{eq-aux:tradeoffpspin} shows that $\gamma$ must fall linearly with $n$ for the optimized algorithm to be noise resilient. In contrast, Fig.~\ref{fig:fig4} in the main text shows that increasing the runtime of an adiabatic algorithm can arbitrarily decrease its fragility, regardless of the system size.

\section{Fragility of expectation values}
\label{app:frag-expectation}

In the main text, we focus on the state-fragility of a quantum algorithm. Here, we consider an alternative figure of merit for noise-resilience that focuses on the perturbation of a cost function $C$.

Consider
\begin{align}
\label{eq:robustnessC} 
\robust^{\langle C \rangle}_Q &\coloneqq  \big(  \bra{\delta \psi_D} C \ket{\delta \psi_D} - \bra{\psi_D} C \ket{\psi_D} \big)^2.
\end{align}
Equation~\eqref{eq:robustnessC} compares the expectation values $ \bra{\psi_D} C \ket{\psi_D}$ and $\bra{\delta \psi_D} C \ket{\delta \psi_D}$ of an observable $C$ under ideal and perturbed implementations of an algorithm, respectively. Since $\ket{\delta \psi_D}$ depends on the noises $Q_l^q$ affecting the system, so does the resilience measure $\robust^{\langle C \rangle}_Q$.

Throughout this Appendix, we take $\robust^{\langle C \rangle}_Q$ as a fragility measure of algorithms that aim to minimize a cost function. This is relevant, for example, for variational and optimization algorithms~\cite{farhi2015quantum, bauer2020quantum, cerezo2021variational}. The implementation of an algorithm is resilient with respect to the cost function $\langle C \rangle \coloneqq \bra{\psi_D} C \ket{\psi_D}$ when $\robust^{\langle C \rangle}_Q \approx 0$.

\subsection{Relating $\robust_Q$ and $\robust^{\langle C \rangle}_Q$}

If an algorithm is state-resilient it is resilient with respect to any cost function $\langle C \rangle$, but the reverse is not true. This is because $\robust_Q$ is a distance between states, and there can be states that are far from each other while an observable $C$ takes the same expectation value~\cite{nielsen2010quantum}. 

Consider a cost function
\begin{equation}\label{Relating_robustness_cost}
    C_{\ket{\phi}} \coloneqq \id-\ketbra{\phi}{\phi}.
\end{equation}
The cost function in Eq.~\eqref{Relating_robustness_cost} could assess the performance of an algorithm to prepare state $\ket{\phi}$. Assume an ideal implementation of an algorithm where the final state is $\ket{\psi_D} = \ket{\phi}$.  Substituting $C_{\ket{\phi}}$ in Eq.~\eqref{eq:robustnessC}, we obtain 
\begin{eqnarray}
    \robust^{\langle C_{\ket{\phi}} \rangle}_Q &=& \Big(\bra{\delta\psi_D}\left(\mathbb{I}-\ketbra{\psi_D}{\psi_D}\right)\ket{\delta\psi_D}-\bra{\psi_D}\left(\mathbb{I}-\ketbra{\psi_D}{\psi_D}\right)\ket{\psi_D}\Big)^2 = \Big(1-|\bra{\delta\psi_D}\psi_D\rangle|^2\Big)^2\nonumber\\
    &=& \Big(1+|\bra{\delta\psi_D}\psi_D\rangle|\Big)^2\Big(1-|\bra{\delta\psi_D}\psi_D\rangle|\Big)^2 = \Big(1+|\bra{\delta\psi_D}\psi_D\rangle|\Big)^2\left(\frac{\robust_Q}{4}\right)^2.
\end{eqnarray}
Then, $\robust^{\langle C_{\ket{\phi}} \rangle}_Q = 0$ implies $\robust_Q = 0$, so both measures of fragility are consistent with each other.

\subsection{Resilience to perturbative noise}

In this section, we derive the expression for the fragility of expectation values, $\robust^{\langle C \rangle}_Q$, of a quantum algorithm under perturbative noise. We prove that, to leading orders in $\delta \theta_j^q$, the following expression quantifies an algorithm's fragility against the perturbative noise process $Q_l^q$:

\begin{align}
\label{eq:ResultRobustnessC}
\robust^{\langle C \rangle}_Q &\approx   \sum_{j,k=1}^D \sum_{q=1}^{\mathcal{N}_j}\sum_{r=1}^{\mathcal{N}_k}   \bra{\psi_0} \left[ Q_j^q(t_j), C(t_D) \right]  \ket{\psi_0}           \bra{\psi_0} \left[ C(t_D),Q_k^r(t_j) \right]  \ket{\psi_0} \delta \theta_j^q \delta \theta_k^r. 
\end{align}
Operators evolving in the Heisenberg picture are denoted by $C(t_j) \nobreak \coloneqq \nobreak \left( \prod_{l=1}^j \bigotimes_{r=1}^{\mathcal{N}_l} M_l^rV_l^r \right)^\dag C \left( \prod_{l=1}^j \bigotimes_{r=1}^{\mathcal{N}_l} M_l^rV_l^r \right)$, where $t_j$ identifies the time elapsed between the start of the protocol and the $j$-th layer of the circuit.
 
We prove it as follows. Under a perturbed dynamics one obtains $\bra{\delta\psi_D} C \ket{\delta\psi_D}$.
A $2$nd order Taylor expansion gives
\begin{align}
\label{eq-app:secondorder}
    \bra{\delta\psi_D} C \ket{\delta\psi_D} \approx \bra{\psi_D} C \ket{\psi_D} &+ \sum_{j=1}^D \sum_{q}^{\mathcal{N}_j}  \left( \frac{\partial}{\partial \delta \theta_j^q}  \bra{\delta\psi_D} C \ket{\delta\psi_D} \right) \at \delta \theta_j^q  \\
    &+ \frac{1}{2} \sum_{j,k=1}^D \sum_{q}^{\mathcal{N}_j}\sum_{r}^{\mathcal{N}_k} \left( \frac{\partial^2}{\partial \, \delta \theta_j^q \delta \theta_k^r}  \bra{\delta\psi_D} C \ket{\delta\psi_D} \right) \at \delta \theta_j^q \delta \theta_k^r.  \nonumber
\end{align}
A derivation similar to that of Eq.~\eqref{eq-app:Statederiv} gives
\begin{align}
\label{eq-app:statederiv2}
\frac{\partial}{\delta\theta_j^q} \ket{\delta\psi_D } &=\prod_{m=j+1}^D \bigotimes_{q'=1}^{\mathcal{N}_m} e^{-i \delta \theta_m^{q'} Q_m^{q'}} M_m^{q'} V_m^{q'} \big( -i Q_j^q \big) \prod_{l=1}^j \bigotimes_{r=1}^{\mathcal{N}_l} e^{-i \delta\theta_l^r Q_l^r} M_l^{r} V_l^r \ket{\psi_0}, \nonumber \\
     \frac{\partial}{\delta\theta_j^q  } \ket{\delta\psi_D } \at &= -i U_{1;L} Q_j^q(t_j)  \ket{\psi_0},
\end{align}
and
\begin{align}
\label{eq-app:statederiv3}
     \frac{\partial^2}{\delta\theta_j^q \delta\theta_k^r} \ket{\delta\psi_D } &= \frac{\partial}{\delta\theta_k^r  } \left( \prod_{m=j+1}^D \bigotimes_{q'=1}^{\mathcal{N}_m}  e^{-i \delta \theta_m^{q'} Q_m^{q'}} \right) M_m^{q'} V_m^{q'} \big( -i Q_j^q \big) \prod_{l=1}^j \bigotimes_{r'=1}^{\mathcal{N}_l} e^{-i \delta\theta_l^{r'} Q_l^{r'}} M_l^{r'} V_l^{r'} \ket{\psi_0} \nonumber \\
     &=\prod_{m''=k+1}^D \bigotimes_{q''=1}^{\mathcal{N}_{m''}}   e^{-i \delta \theta_{m''}^{q''} Q_{m''}^{q''}} M_{m''}^{q''} V_{m''}^{q''} \left( -i Q_k^r \right)  \prod_{m=j+1}^k \bigotimes_{q'=1}^{\mathcal{N}_m}  e^{-i \delta \theta_m^{q'} Q_m^{q'}} M_m^{q'} V_m^{q'} \big( -i Q_j^q \big)  \nonumber \\
     & \qquad \qquad \qquad \qquad \qquad \qquad \qquad \prod_{l=1}^j \bigotimes_{r'=1}^{\mathcal{N}_l} e^{-i \delta\theta_l^{r'} Q_l^{r'}} M_l^{r'} V_l^{r'} \ket{\psi_0}, \nonumber \\
    \frac{\partial^2}{\delta\theta_j^q \delta\theta_k^r  } \ket{\delta\psi_D } \at &= - U_{k+1;L} Q_k^r U_{j+1;k} Q_j^q U_{1;j} \ket{\psi_0} = - U_{k+1;L} Q_k^r U_{1;k} Q_j^q(t_j)  \ket{\psi_0} \nonumber \\
    &= - U_{k+1;L} U_{1;k} U_{1;k}^\dag Q_k^r U_{1;k} Q_j^q(t_j)  \ket{\psi_0} = - U_{1;L}   Q_k^r(t_k)  Q_j^q(t_j)  \ket{\psi_0},
\end{align}
where we assumed $k \geq j$ without loss of generality. By inserting Eqs.~\eqref{eq-app:statederiv2} and~\eqref{eq-app:statederiv3} into Eq.~\eqref{eq-app:secondorder} we obtain 
\begin{align}
\label{eq-app:secondorderV2}
     \bra{\delta\psi_D} C \ket{\delta\psi_D} &\approx \bra{\psi_D} C \ket{\psi_D} + \sum_{j=1}^D \sum_{q}^{\mathcal{N}_j}  \left[ \left( \frac{\partial}{\partial \delta \theta_j^q}  \bra{\delta\psi_D} \right) C \ket{\delta\psi_D} +   \bra{\delta\psi_D} C \left( \frac{\partial}{\partial \delta \theta_j^q} \ket{\delta\psi_D} \right) \right] \at \delta \theta_j^q \nonumber \\
     &+ \frac{1}{2} \sum_{j,k=1}^D \sum_{q}^{\mathcal{N}_j}\sum_{r}^{\mathcal{N}_k} \bigg[ \left( \frac{\partial^2}{\partial \delta \theta_j^q \delta \theta_k^r}  \bra{\delta\psi_D} \right) C \ket{\delta\psi_D} +  \bra{\delta\psi_D} C \left( \frac{\partial^2}{\partial \delta \theta_j^q \delta \theta_k^r}  \ket{\delta\psi_D} \right)  \nonumber \\
     &\qquad \qquad \qquad \qquad + \left( \frac{\partial}{\delta \theta_j^q}  \bra{\delta\psi_D} \right) C \left( \frac{\partial}{\partial \delta \theta_k^r}  \ket{\delta\psi_D} \right) + \left( \frac{\partial}{\partial \delta \theta_k^r}  \bra{\delta\psi_D} \right) C \left( \frac{\partial}{\partial \delta \theta_j^q}  \ket{\delta\psi_D} \right) \bigg] \at \delta \theta_j^q \delta \theta_k^r \nonumber \\
    &= \bra{\psi_D} C \ket{\psi_D} + \sum_{j=1}^D \sum_{q}^{\mathcal{N}_j}  \bigg[ +i \bra{\psi_0} Q_j^q(t_j) U_{1;L}^\dag C U_{1;L} \ket{\psi_0} - i \bra{\psi_0} U_{1;L}^\dag C   U_{1;L} Q_j^q(t_j)  \ket{\psi_0}   \bigg]  \delta \theta_j^q \nonumber \\
    &+ \frac{1}{2} \sum_{j,k=1}^D \sum_{q}^{\mathcal{N}_j}\sum_{r}^{\mathcal{N}_k} \bigg[ -\bra{\psi_0} Q_j^q(t_j) Q_k^r(t_k) U_{1;L}^\dag C U_{1;L} \ket{\psi_0} -  \bra{\psi_0} U_{1;L}^\dag C U_{1;L} Q_k^r(t_k) Q_j^q(t_j)  \ket{\psi_0}   \nonumber \\
     &\qquad \qquad \qquad + \bra{\psi_0} Q_j^q(t_j) U_{1;L}^\dag C U_{1;L} Q_k^r(t_k) \ket{\psi_0} + \bra{\psi_0} Q_k^r(t_k) U_{1;L}^\dag C U_{1;L} Q_j^q(t_j) \ket{\psi_0} \bigg]  \delta \theta_j^q \delta \theta_k^r \nonumber \\ 
     &= \bra{\psi_D} C \ket{\psi_D} + \sum_{j=1}^D \sum_{q}^{\mathcal{N}_j}  \bigg[ +i \bra{\psi_0} Q_j^q(t_j) C(t_D)  \ket{\psi_0} - i \bra{\psi_0} C(t_D) Q_j^q(t_j)  \ket{\psi_0}   \bigg]  \delta \theta_j^q \nonumber \\
    &+ \frac{1}{2} \sum_{j,k=1}^D \sum_{q}^{\mathcal{N}_j}\sum_{r}^{\mathcal{N}_k} \bigg[ -\bra{\psi_0} Q_j^q(t_j) Q_k^r(t_k) C(t_D) \ket{\psi_0} -  \bra{\psi_0} C(t_D) Q_k^r(t_k) Q_j^q(t_j)  \ket{\psi_0}   \nonumber \\
     &\qquad \qquad \qquad + \bra{\psi_0} Q_j^q(t_j) C(t_D) Q_k^r(t_k) \ket{\psi_0} + \bra{\psi_0} Q_k^r(t_k) C(t_D) Q_j^q(t_j) \ket{\psi_0} \bigg]  \delta \theta_j^q \delta \theta_k^r \nonumber \\
     &= \bra{\psi_D} C \ket{\psi_D} + \sum_{j=1}^D \sum_{q}^{\mathcal{N}_j}   i \bra{\psi_0} \left[ Q_j^q(t_j), C(t_D) \right]  \ket{\psi_0}   \delta \theta_j^q \nonumber \\
    &+ \frac{1}{2} \sum_{j,k=1}^D \sum_{q}^{\mathcal{N}_j}\sum_{r}^{\mathcal{N}_k} \bigg( \bra{\psi_0} Q_j^q(t_j) \left[ C(t_D), Q_k^r(t_k) \right] \ket{\psi_0} -  \bra{\psi_0} \left[ C(t_D) , Q_k^r(t_k) \right] Q_j^q(t_j)  \ket{\psi_0} \bigg)  \delta \theta_j^q \delta \theta_k^r.
    \end{align}
Thus
\begin{align} 
   \bra{\delta \psi_D} C \ket{\delta \psi_D} - \bra{\psi_D} C \ket{\psi_D}
   &\approx   \sum_{j=1}^D \sum_{q}^{\mathcal{N}_j}   i \bra{\psi_0} \left[ Q_j^q(t_j), C(t_D) \right]  \ket{\psi_0}   \delta \theta_j^q   \nonumber \\
   &  +     \frac{1}{2} \sum_{j,k=1}^D \sum_{q}^{\mathcal{N}_j}\sum_{r}^{\mathcal{N}_k}  \bra{\psi_0} \left[ Q_j^q(t_j), \left[ C(t_D), Q_k^r(t_k) \right] \right] \ket{\psi_0}   \delta \theta_j^q \delta \theta_k^r.
\end{align}
The leading order in the cost function's fragility is
\begin{align}
    \robust^{\langle C \rangle}_Q &\coloneqq 
 \Big(  \bra{\delta \psi_D} C \ket{\delta \psi_D} - \bra{\psi_D} C \ket{\psi_D} \Big)^2 \approx \sum_{j,k=1}^D \sum_{q}^{\mathcal{N}_j}\sum_{r}^{\mathcal{N}_k}   \bra{\psi_0} \left[ Q_j^q(t_j), C(t_D) \right]  \ket{\psi_0} \bra{\psi_0} \left[ C(t_D) , Q_k^r(t_j) \right]  \ket{\psi_0}   \delta \theta_j^q  \delta \theta_k^r,
\end{align}
which proves Eq.~\eqref{eq:ResultRobustnessC}. 

Note that the leading term is null if the unperturbed algorithm reaches an extremum of $\langle C \rangle$. This is because the terms $\bra{\psi_0} \left[ Q_j^q(t_j), C(t_D) \right]  \ket{\psi_0}$ can be interpreted as derivatives of $\langle C(t_D) \rangle$, which are null at an extremum of $\langle C \rangle$. In this case, the leading non-zero term in $\robust^{\langle C \rangle}_Q$ is of order $\mathcal{O} (\delta \theta^4)$.

\subsection{Resilience to uncorrelated noise}

By averaging Eq.~\eqref{eq:ResultRobustness} over the noises $\delta\theta_j^q$, we obtain expressions that characterize the fragility of an algorithm's implementation against uncorrelated noises:
\begin{align}
\label{eq:ResultMeanRobustnessC}
\overline{\robust^{\langle C \rangle}_Q} &\approx \sum_{l=1}^D \sum_{q=1}^{\mathcal{N}_l} \sigma_{lq}^2 \, \left| \bra{\psi_0} \Big[ Q_l^q(t_l), C(t_D) \Big] \ket{\psi_0} \right|^2   . 
\end{align}

The time-continuous version is
\begin{align}
\label{eq:meanResultRobustnessCCont}
\overline{\robust^{\langle C \rangle}_Q } &\approx   \int_0^T   \gamma_t \Big| \bra{\psi_t} \left[ Q_t, C(T) \right]  \ket{\psi_t} \Big|^2 \, dt, 
\end{align}
where $T$ is the algorithm's runtime and $\gamma_t$ characterizes the noise's intensity, with $\overline{\xi_t \xi_{t'}} = \delta(t-t') \gamma_t$.
 
\subsection{Resilience-runtime tradeoff relations for cost functions}

A tradeoff relation analogous to Eq.~\eqref{eq:tradeoffState} in the main text constrains the resilience of expectation values and the number of error gates:
\begin{align}
\label{eq:tradeoffCost}
     N_G \, \overline{\robust^Q_{\langle C \rangle}} \geq  \left( \sum_{\textnormal{path}} |d_Q \langle C \rangle|  \right)^2.
\end{align}
The right-hand side is
\begin{align}
    \sum_{\textnormal{path}}  |d_Q\langle C \rangle|  \coloneqq \sum_{l q} \sigma_{l}^{q} \Big| -i \bra{\psi_0} \big[Q_l^q(t_l), C(t_D)\big] \ket{\psi_0}\Big|.
\end{align}
Each term in the sum corresponds to the (first order) change in $\langle C \rangle$ if the system were driven by unitary dynamics $e^{-i \sigma_{lq}Q_l^q}$.

\end{document}